\documentclass[a4paper]{article}
\usepackage{geometry}
\usepackage{graphicx}
\usepackage{amssymb}
\usepackage{url,hyperref}
\usepackage{epstopdf}
\usepackage{verbatim}
\usepackage{calc,array}
\usepackage{rotating}
\usepackage{booktabs}
\usepackage{blindtext}
\usepackage{fancybox}
\usepackage[toc,page]{appendix}
\usepackage{amsmath} 
\usepackage{xcolor}
\usepackage{authblk}

\renewcommand{\vec}[1]{\mathbf{#1}}
\newcommand{\sigmaHH}{\sigma_{\rm HH}}
\newcommand{\sigmaHHMC}{\sigma_{\rm HH, MC}}
\newcommand{\sigmaHHSMNNLO}{\sigma_{\rm HH, NNLO+NNLL}^{\rm SM}}

\newcommand{\RHH}{R_{\rm HH}}
\newcommand{\mH}{m_{\rm H}}
\newcommand{\pTH}{p_{\rm T, H}}
\newcommand{\PH}{\rm H}
\newcommand{\mHH}{m_{\rm HH}}
\newcommand{\HH}{\rm HH}
\def\@preprint{\@empty}
\newcommand\preprint[1]{\gdef\@preprint{\hfill }}

\usepackage[style=numeric, 
            backref=true,
            style=numeric-comp,
            firstinits=true, 
            maxcitenames=300,
            maxbibnames=500,
            block=none,
            sortcites=true,  
            sorting=none,
            backend=bibtex
            ]{biblatex}
\title{\boldmath Analytical parametrization and shape classification of anomalous HH production in the EFT approach - LHCHXSWG-2016-001}
\author[1]{Alexandra Carvalho}
\author[1]{Martino Dall'Osso}
\author[1]{Pablo de Castro Manzano}
\author[1]{Tommaso Dorigo}
\author[2]{Florian Goertz}
\author[3]{Maxime Gouzevich}
\author[2]{Mia Tosi}

\affil[1]{Dipartimento di Fisica e Astronomia and INFN, Sezione di Padova, Via Marzolo 8, I-35131 Padova, Italy. }
\affil[2]{ CERN, 1211 Geneva 23, Switzerland }
\affil[3]{ Universite de Lyon, Universie Claude Bernard Lyon 1, CNRS-IN2P3, Institut de
Physique Nucleaire de Lyon, Villeurbanne, France}

\begin{document}
\maketitle
\begin{abstract}
\noindent In this document we study the effect of anomalous 
Higgs boson couplings on non-resonant pair production of Higgs bosons ($\HH$) at the LHC. 
We explore the space of the five parameters $\kappa_{\lambda}$, $\kappa_{t}$,  
$c_2$, $c_g$, and $c_{2g}$ in terms of the corresponding kinematics of the
final state, and describe a partition of the space into a limited number of 
regions featuring similar phenomenology in the kinematics of $\HH$ final state. We call {\em clusters} the sets of points belonging to the same region; to each cluster corresponds a representative point which we call a {\em benchmark}.  
We discuss a possible technique to estimate the sensitivity of an experimental search to the kinematical differences between the phenomenology of the benchmark points and the rest of the parameter space contained in the corresponding cluster. 
We also provide an analytical 
parametrization of the cross-section modifications that the variation of anomalous
couplings produces with respect to standard model $\HH$ production along with a recipe to translate the results into other parameter-space
bases. Finally, we provide a preliminary analysis of variations in the topology of the final state within each region based on recent LHC results.
\end{abstract}

\tableofcontents

\clearpage
\section{Introduction}

The present work stems from the studies we have undertaken to attempt an exhaustive
description of the complex parameter space that describes the possible
modifications of standard model (SM) production of Higgs boson ($\PH$) pairs produced by anomalous
couplings. The characterization of the phenomenology may be done by considering the shape 
of density functions of kinematic pseudo-observables fully specifying the production process. 
By considering kinematical quantities describing $\HH$ production at Leading
Order (LO), without the inclusion of any initial- or final-state effects nor the decay 
of the Higgs bosons, one may concentrate on the similarities and the differences produced by
distinct physics scenarios, determined by the value of the five anomalous coupling
parameters.

While the previous work~\cite{Dall'Osso:2015aia} focused on the qualitative 
taxonomy of the kinematics induced by di-Higgs production, in this
paper we consider mainly the cross section of that process, offering an
useful analytical parametrization. In section~\ref{sec:bench} we provide the parametrization of the Lagrangian density in 
terms of five anomalous coupling parameters.  In section~\ref{sec:shape} we recall the
results of our clustering procedure, and discuss the properties of the identified benchmarks and the intra-cluster variability. In particular the clustering procedure is compared to the first experimental results from the LHC.
In section \ref{sec:cx} we derive the analytical parametrization of the cross section, discuss its precision, and offer a recasting recipe to use the formula in other bases.

\section{Higgs boson pair production by gluon-gluon fusion}
\label{sec:bench}

In the context of Beyond the Standard Model (BSM) theories, di-Higgs production in gluon-gluon fusion can be 
described to leading approximation with the Lagrangian~\cite{Buchalla:2137956}
\begin{equation}
\begin{split}
{\cal L}_{\PH} =  &\frac{1}{2} \partial_{\mu}\, \PH \partial^{\mu} \PH - \frac{1}{2} m_{\PH}^2 \PH^2 -  {\kappa_{\lambda}}\,  \lambda_{SM} \nu\, \PH^3 \\
 & - \frac{ m_t}{v}(v+   {\kappa_t} \,   \PH  +  \frac{c_{2}}{\nu}   \, \HH ) \,( \bar{t_L}t_R + h.c.)  + \frac{1}{4} \frac{\alpha_s}{3 \pi \nu} (   c_g \, \PH -  \frac{c_{2g}}{2 \nu} \, \HH ) \,  G^{\mu \nu}G_{\mu\nu}\,,
\end{split}
\label{eq:lag}
\end{equation}
\noindent
where $\nu = 246\,$GeV is the vacuum expectation value of the Higgs field.
This Lagrangian includes five parameters: the deviation of the Higgs boson trilinear coupling
$\lambda$ (top Yukawa coupling $y_t$) from its SM value $\lambda_{SM}$
($y_{t, SM} = \sqrt 2 m_t/v$) quantified by $\kappa_{\lambda} \equiv \lambda/\lambda_{SM}$
($\kappa_t \equiv y_t/y_{t}^{\rm SM}$), as well as the coefficients of three pure BSM operators which describe
the contact interaction between two Higgs bosons and two top quarks ($c_2$),
and the Higgs boson contact interaction with one  ($c_g$) and two gluons ($c_{2g}$).
In the Effective Field Theory (EFT) description, modifications of
the interactions between the Higgs boson and the other SM fields are generated by 
higher-dimensional operators, which after electroweak symmetry breaking induce the couplings
above. While assuming a linear realization of the SM gauge symmetry, {\it i.e.}, 
assuming the H boson as part of a weak doublet, leads to relations between these couplings
 (in the case of a linear realization with dimension-6 operators, we get $c_{2g} = − c_g$, see
Ref.~\cite{Buchmuller:1985jz}), those are lifted in the non-linear realization. We do not consider a possible enhanced 
coupling of the Higgs boson with bottom quarks, which are already constrained experimentally~\cite{Khachatryan:2016vau}.

\begin{figure}[h]
\centering
\includegraphics[scale=0.75]{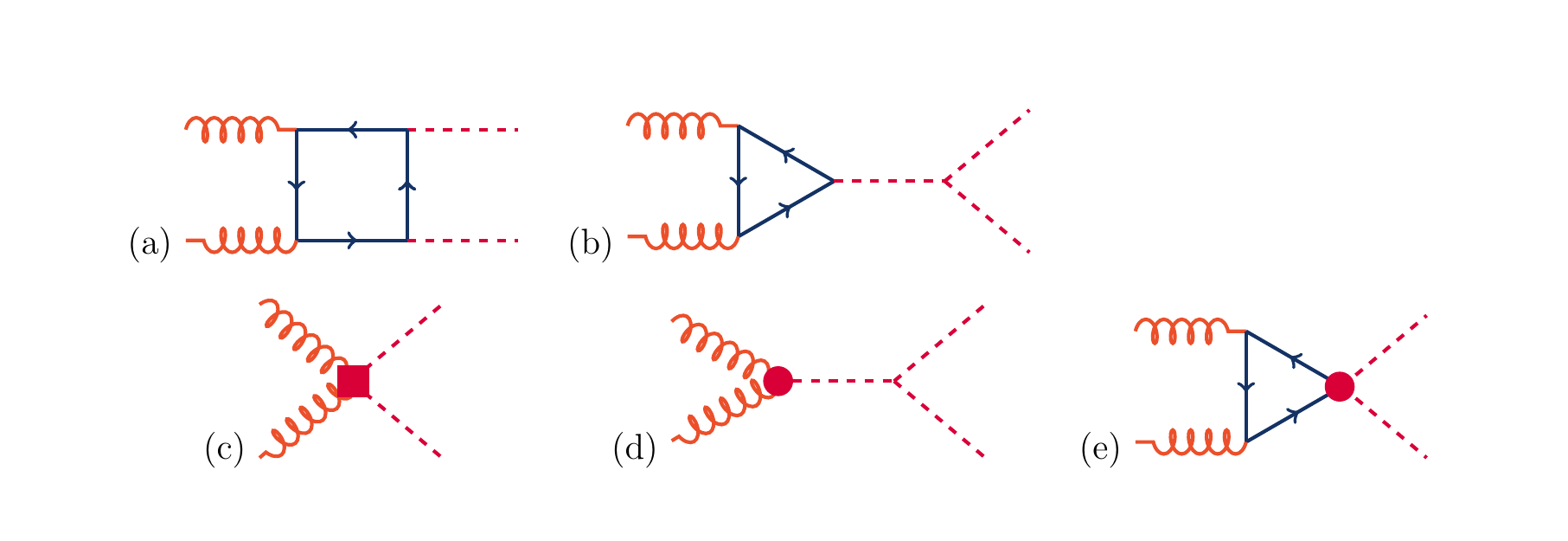}
\caption{\small Generic Feynman diagrams that contribute to $\HH$ production via gluon-gluon fusion at leading order. Diagrams (a) and (b) correspond to SM-like processes, while diagrams (c), (d) and (e) include pure BSM vertices: (c) and (d) describe contact interactions between the $\PH$ boson and gluons, and (e) exploits the contact interaction of two $\PH$ bosons with top quarks.   \label{fig:dia}}
\end{figure}

Five Feynman diagrams can be constructed from the above Lagrangian (see Fig.~\ref{fig:dia}), 
each corresponding to a matrix element associated
to different combinations of BSM and SM-like parameters and different properties of
the $\HH$ final state.

\section{Kinematic clustering}
\label{sec:shape}

\subsection{Clustering procedure}
\label{sec:clus}

An important consequence of the introduction of the five BSM parameters in the Lagrangian provided in Eq.~(\ref{eq:lag}), in addition to the modification of the overall Higgs boson pair-production cross section, is the generation of significant modifications of the kinematic properties of the final state with respect to the pure SM process. The experimental exploration of the five-dimensional model space is by no means trivial, as the optimization of the search strategy for the $\HH$ signal depends significantly on the investigated parameter space point. 
In order to address this problem, we designed a clustering procedure to group regions of parameter space which can be probed by the same search. In the spirit of this approach, analyses would optimize their selection for a model (a benchmark) chosen to represent at best the kinematic characteristics of the corresponding region. The benchmarks resulting from the clustering procedure are summarized in this note; for implementation details the reader is referred to  Ref.~\cite{Dall'Osso:2015aia}. We recall below the basic ideas of our procedure.

The  $gg \rightarrow \HH$ production is a $2 \rightarrow 2$ process. 
In the center-of-mass reference frame the kinematic properties of the final state can be fully characterized by two variables: the invariant mass of the two Higgs bosons $\mHH$ and the polar angle of one of the bosons in the center-of-momentum frame $|\cos \theta^{*}|$. At leading order, all other observables describing the final state have no connection to the structure of the Lagrangian density from Eq.~\ref{eq:lag}.
The kinematic properties of each EFT point can then be characterized by
estimating the two-dimensional probability density function of $\mHH$ and
$|\cos \theta^{*}|$. By using a suitable test statistic sensitive to
the shape differences of the two-dimensional density function, one may quantify
the kinematic differences between different model points in the
five-dimensional phase-space. The value of the test statistic may then be used
to group the models that are kinematically most similar.
This procedure is referred to as \textit{cluster analysis} and a group of models as \textit{a cluster}. The employed test statistic was a likelihood ratio based on Poisson counts in $\mHH-|\cos \theta^{*}|$ histograms; the clustering was performed through a hierarchical agglomerative technique.

The clustering procedure was applied to a set of models corresponding to a fine
scan in several parameter space directions. The investigated range of the parameters was decided by taking into account the current experimental constraints. In particular, the parameters were allowed to
vary as $|\kappa_{\lambda}| \leq$ 15,  $\kappa_t\subset [0.5,2.5]$,  $|c_2| \leq$3 and $(|c_g|,|c_{2g}|) \leq 1$, where $\kappa_t$ and $c_2$ feature a step size of O(0.5) and $\kappa_{\lambda}$ is varied in O(1) steps.
The granularity of the scan in $c_g$ and $c_{2g}$ is 0.2. To have a better accuracy
in the points of minimal di-Higgs production
cross section, where the changes in kinematics are particularly strong, we increased the density of scanned points in the corresponding regions. 
This resulted in a total of 1507 inspected points, distributed with a variable binning within the boundaries quoted above. The
simulations were performed with the Madgraph\_aMC@NLO version 2.2.1 Monte Carlo (MC) simulation package~\cite{Alwall:2014hca}, using the model provided
by the authors of~\cite{Hespel:2014sla}, were the loop factors including the full $m_{t}$
dependence are calculated on an event-by-event basis with a Fortran routine. The PDF set
used was NNPDF23LO1~\cite{Ball:2012cx} and the factorization and renormalization scale considered were
$\sqrt{\hat{s}}= \mHH$. The relevant input masses were $\mH$ = 126\,GeV
and $m_t$ = 173.18\,GeV.

The cluster analysis leads to the division of the 1507 inspected points into 12 groups. Each group displays similar kinematic characteristics, and clear differences compared with members of the other groups. For each cluster a benchmark is defined as the sample most similar to all the others in the cluster, where the similarity metric is the one given by the test statistic. The numerical details are provided later in sub-section~\ref{clus_res}.

\subsection{Outliers}
\label{sec:intra_outliers}

The benchmarks are chosen to capture well the main features of the cluster kinematics. Nevertheless some of the cluster members
still exhibit residual differences with respect to the benchmark. 
These intra-cluster differences could lead to limited deviations
in the experimental signal efficiencies. If the analysis has a sufficient resolution to resolve those differences we propose here a simple approach to select six extreme cases (referred to as outliers) within each
cluster, which can tentatively be used to evaluate the possible variation of experimental efficiencies within a cluster. If the analyzers want to preserve the simplifications offered by the cluster approach we recommend to fully simulate (generate and propagate through the experimental apparatus) only the benchmark and obtain the outliers through an event-by-event reweighing procedure in the $\mHH-|\cos \theta^{*}|$ space. The results (for example limits) shall be then presented for the benchmark and benchmark reweighted to the outliers. The weights may be easily estimated at generator level.

\begin{figure*}[h]\begin{center}
\includegraphics[width=0.41\textwidth, angle =0 ]{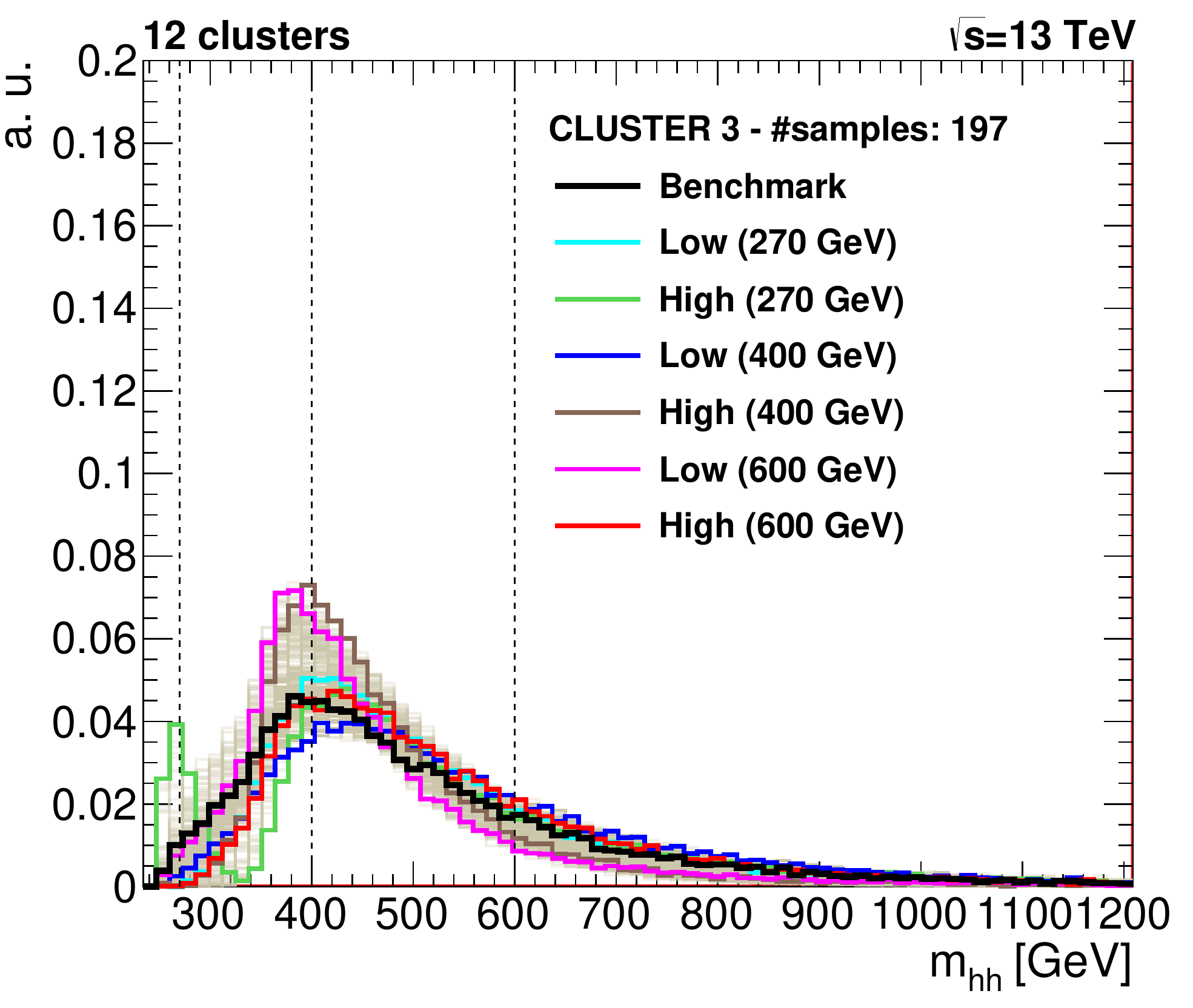}
\includegraphics[width=0.41\textwidth, angle =0 ]{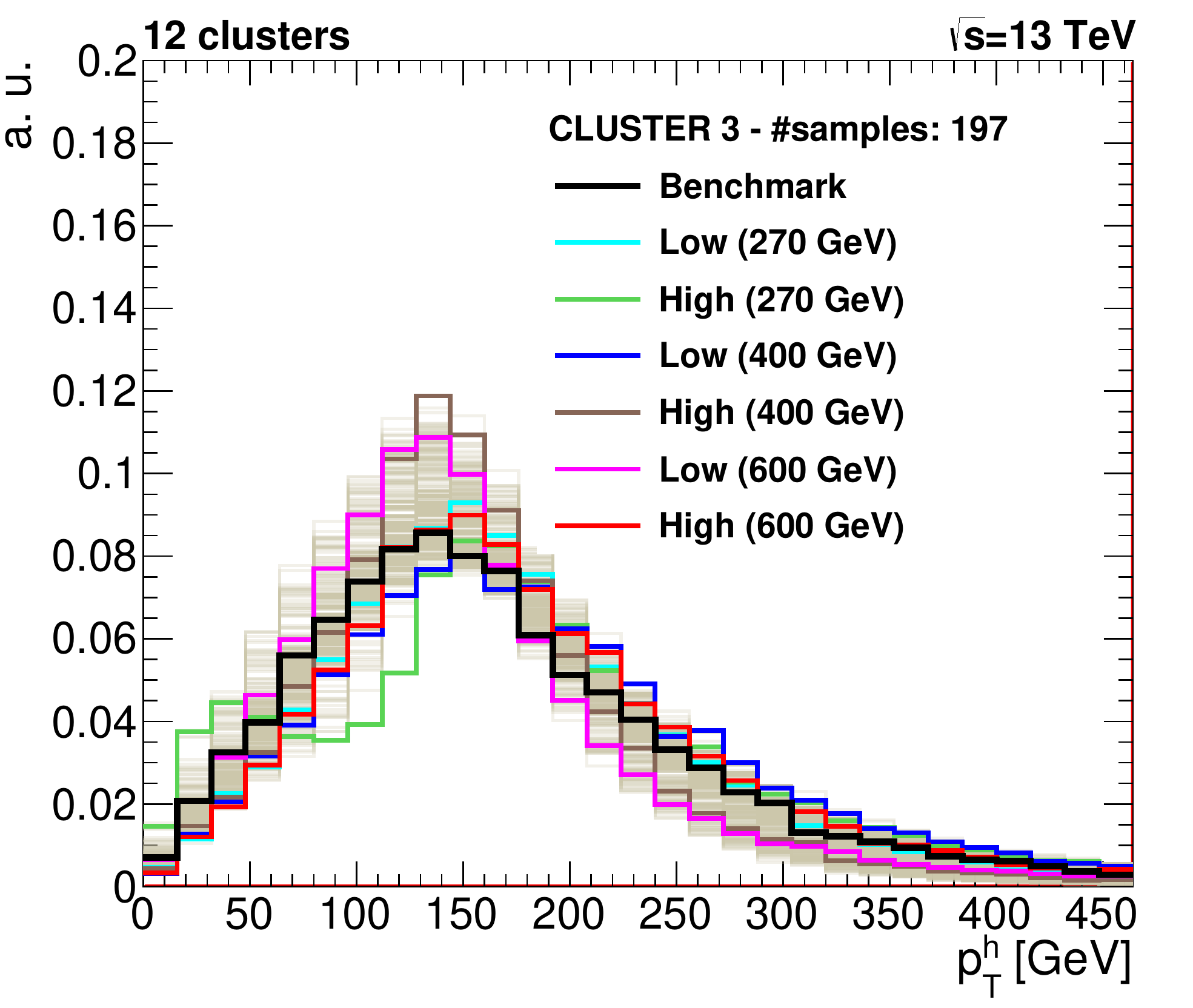}
\includegraphics[width=0.41\textwidth, angle =0 ]{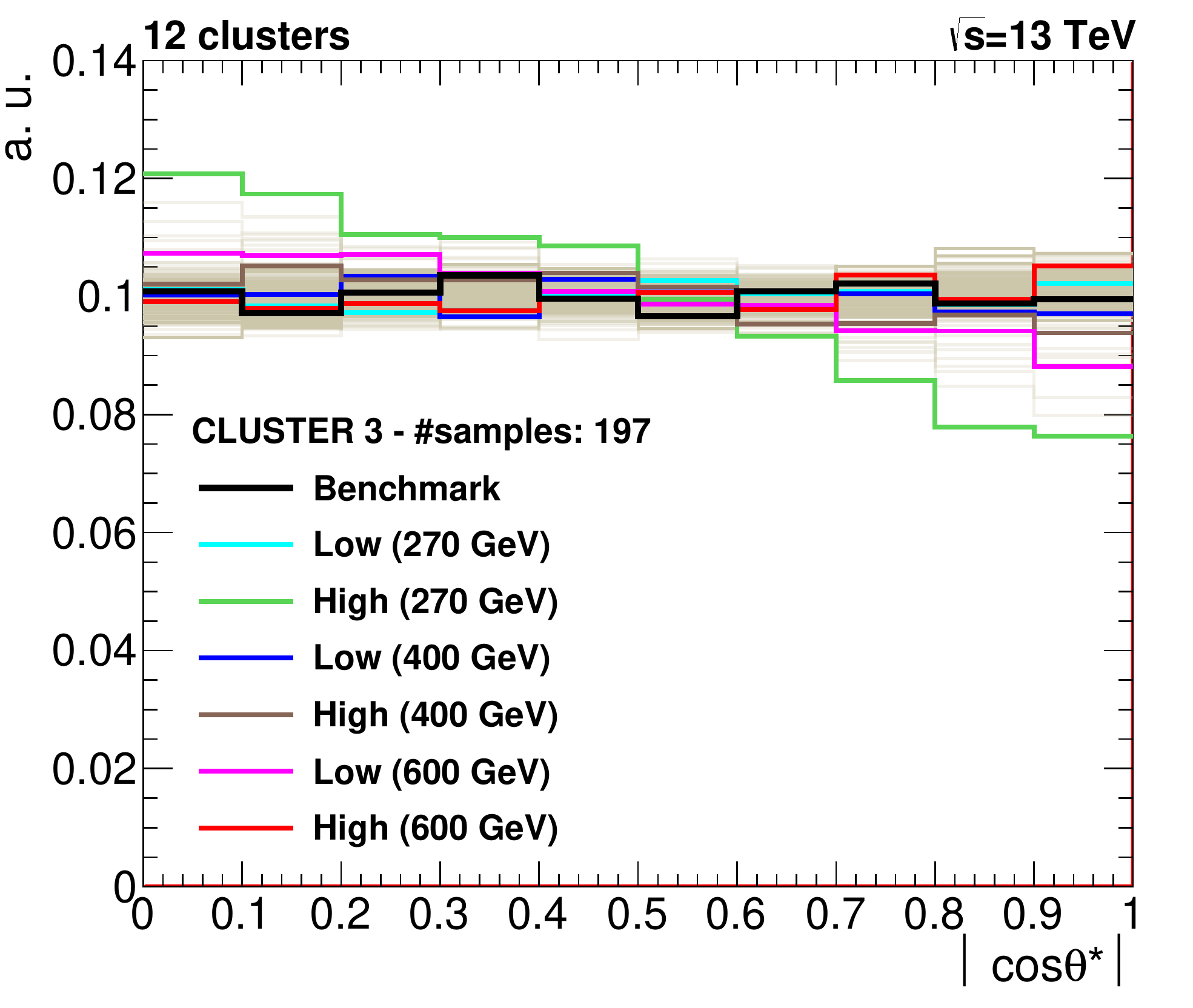}
\caption{\small  The $\mHH$ (top left), $\pTH$ (top right) and $|\cos \theta^{*}|$ (bottom) distributions for the members of cluster 3. The benchmark (in black color) and corresponding outliers (colored lines) are highlighted. The three mass regions are indicated by vertical dashed lines.\label{fig:out_mass}}
\end{center}\end{figure*}

The $\mHH$ distribution features the largest intra-cluster
variation and it is likely the most important distribution
for experimental analyses. We define therefore the outliers as subset of samples that envelope
all the other samples of the cluster in three mass points\footnote{Note that the chosen outliers are susceptible to a different level of
arbitrariness: the choice of the points of the parameter space scan,
the choice of the variable and the points where to search for the cluster
envelope and the histogram binning.} $m_{\rm HH\,,1}\,\equiv$ 270\,GeV,
$m_{\rm HH\,,2}\,\equiv$ 400\,GeV and $m_{\rm HH\,,3}\,\equiv$ 600\,GeV, applied to a histogram with 20\,GeV
wide bins. The first and last mass points are intended to catch
the analysis sensitivity to threshold region ($\mHH \approx 2 m_{\PH}$) and energy-tail modifications. The
intermediate mass point is close to the typical valley found
in the distribution due to a cancellation between the different diagrams, and it is intended
to catch the analyses sensitivity to short-distance fluctuations in shape.
Figure \ref{fig:out_mass} (top) shows an example of the di-Higgs mass distribution for cluster 3 with the outliers.
In Fig.~\ref{fig:out_mass} (bottom) we provide the $\pTH$ and $|\cos \theta^{*}|$ spectra with the same outliers.
One may observe that the choice of outliers in $\mHH$ is also reasonably valid for the two other distributions.

\subsection{Results}
\label{clus_res}

\begin{figure}[hbt]\begin{center}
\includegraphics[width=0.32\textwidth, angle =0 ]{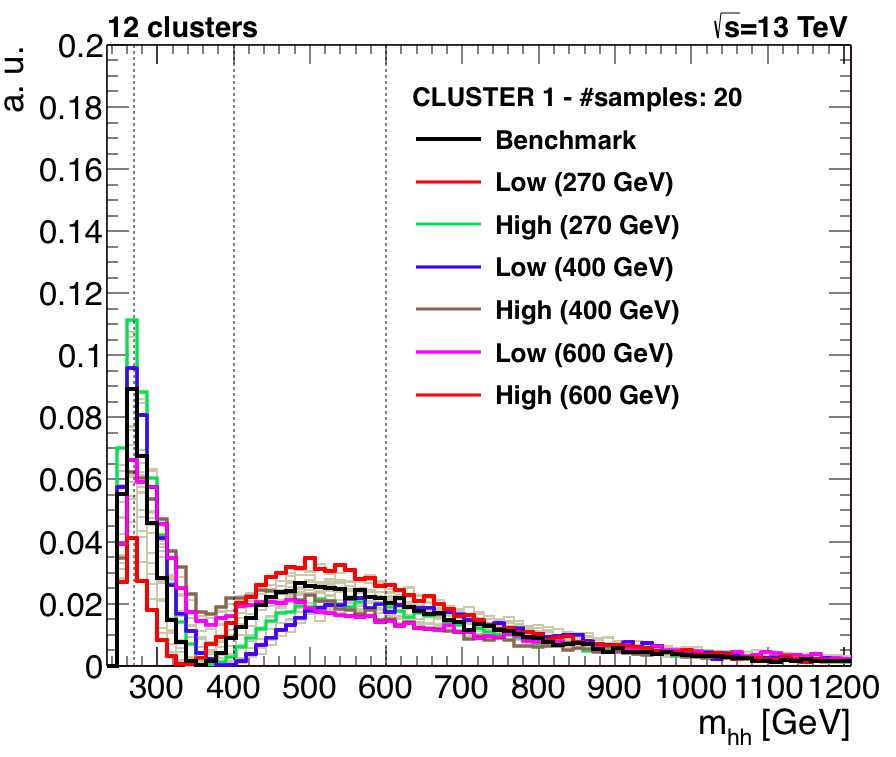}
\includegraphics[width=0.32\textwidth, angle =0 ]{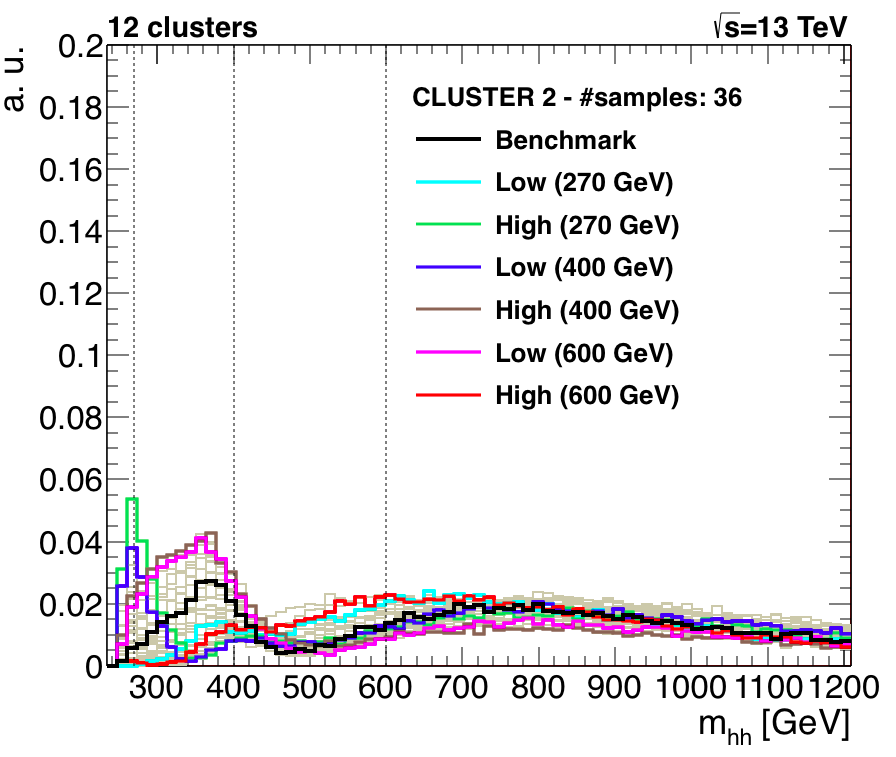}
\includegraphics[width=0.32\textwidth, angle =0 ]{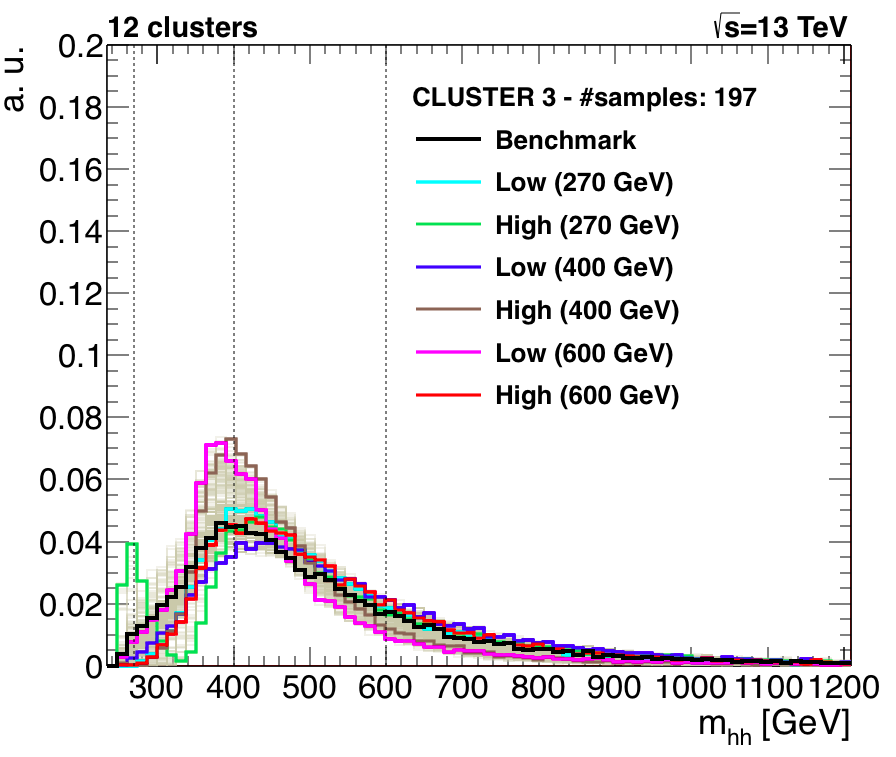}
\includegraphics[width=0.32\textwidth, angle =0 ]{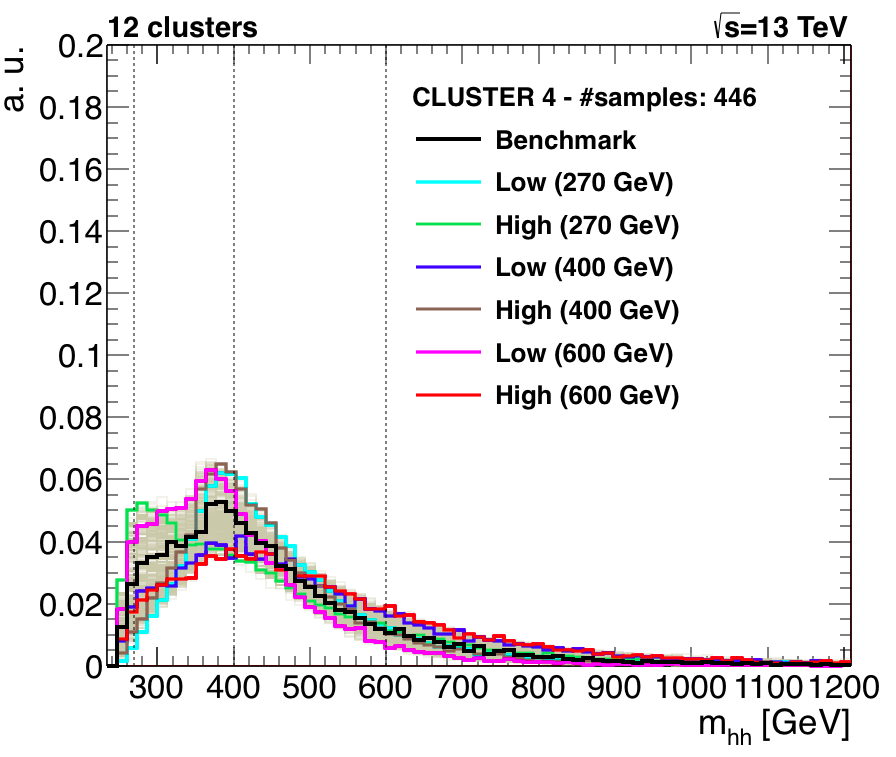}
\includegraphics[width=0.32\textwidth, angle =0 ]{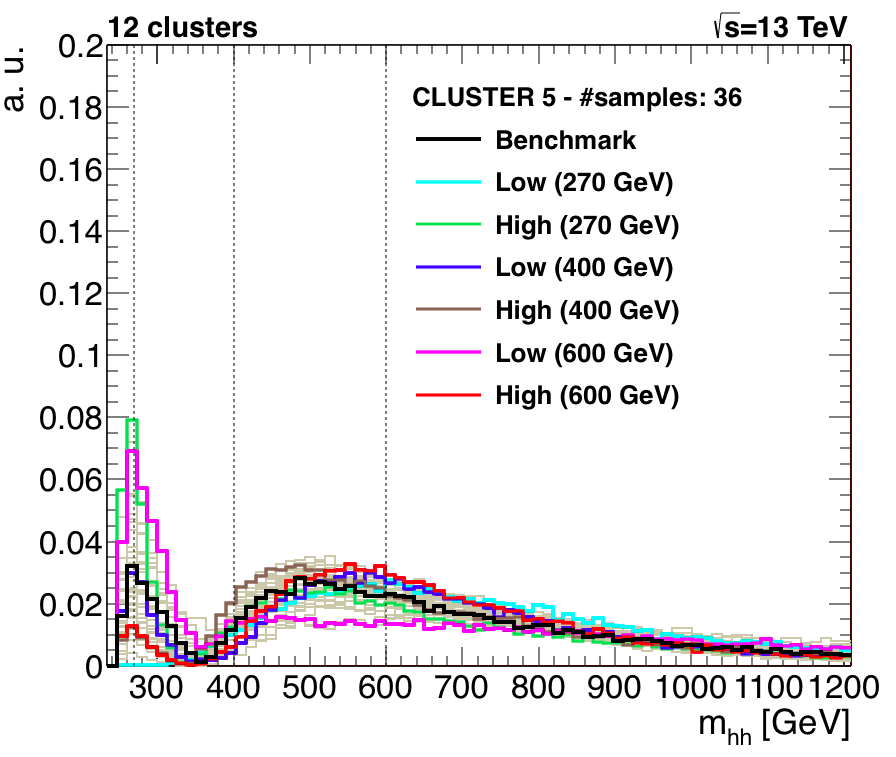}
\includegraphics[width=0.32\textwidth, angle =0 ]{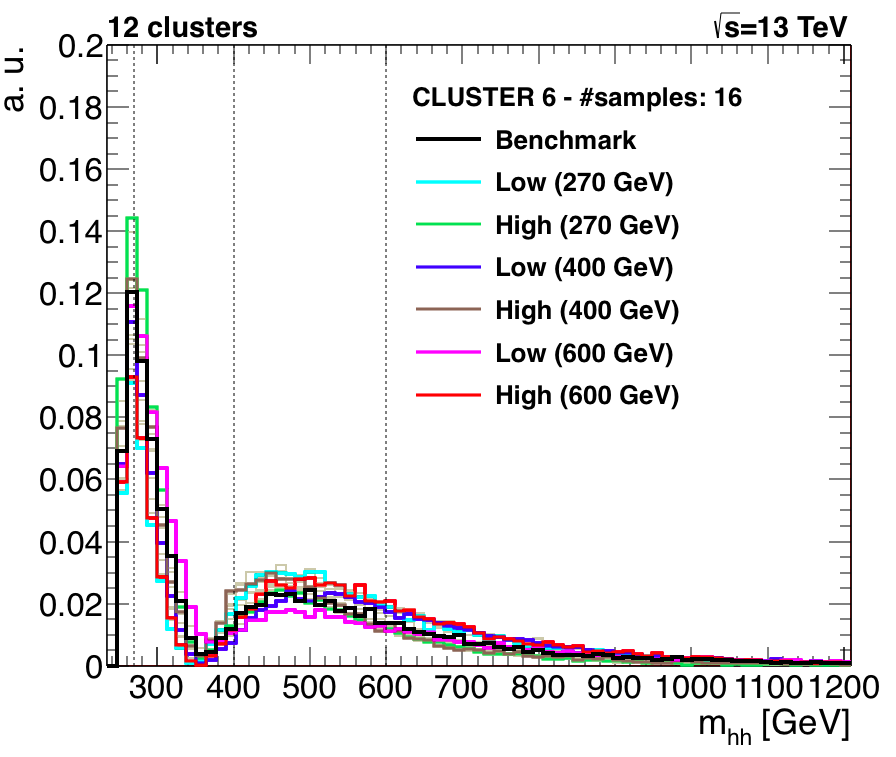}
\includegraphics[width=0.32\textwidth, angle =0 ]{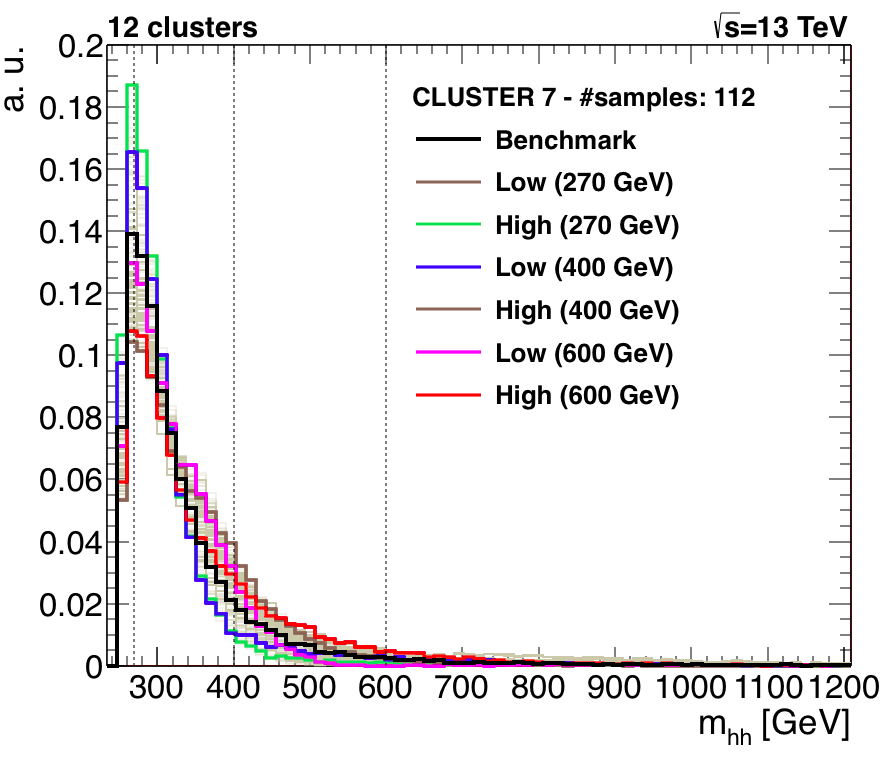}
\includegraphics[width=0.32\textwidth, angle =0 ]{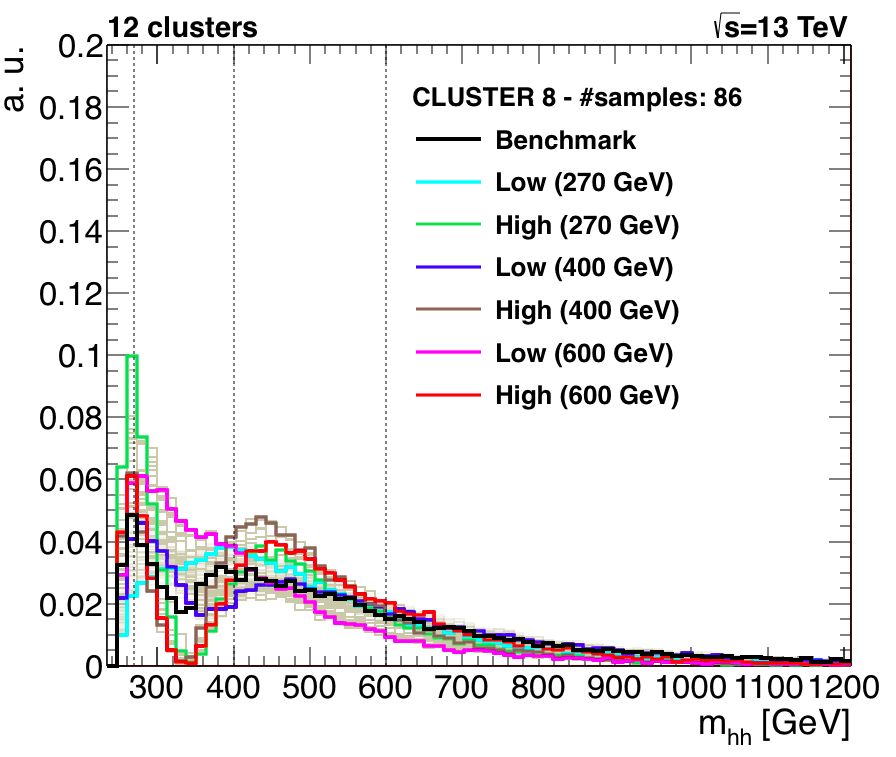}
\includegraphics[width=0.32\textwidth, angle =0 ]{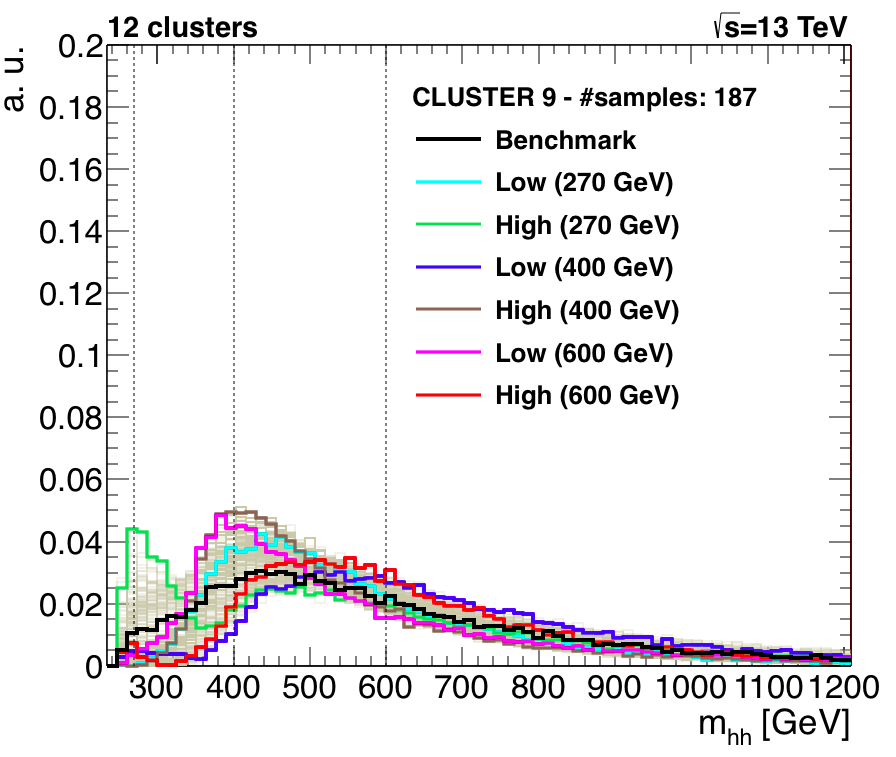}
\includegraphics[width=0.32\textwidth, angle =0 ]{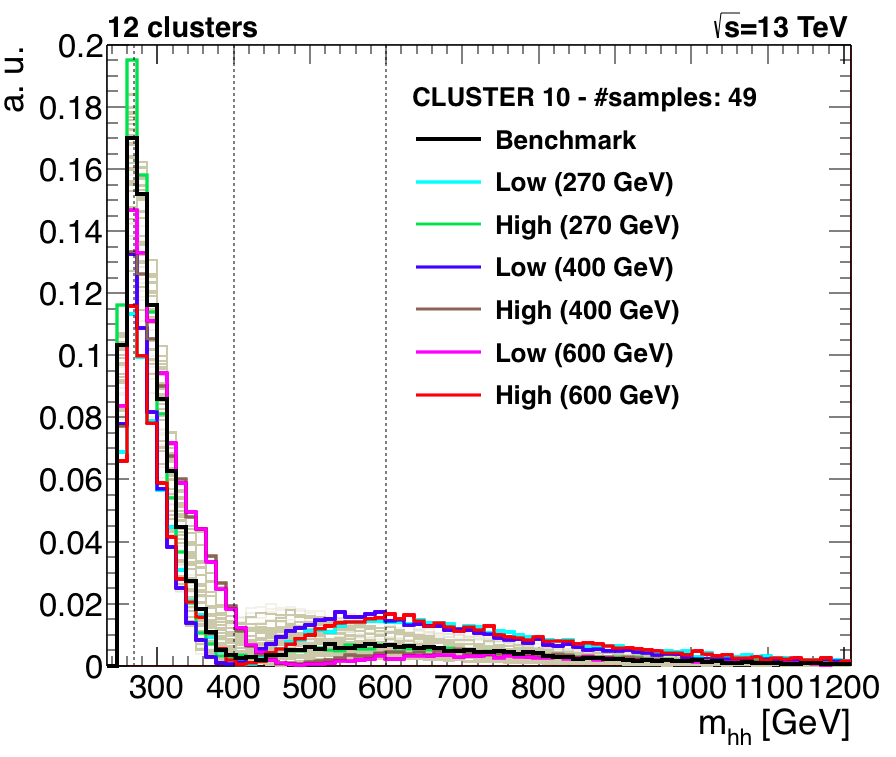}
\includegraphics[width=0.32\textwidth, angle =0 ]{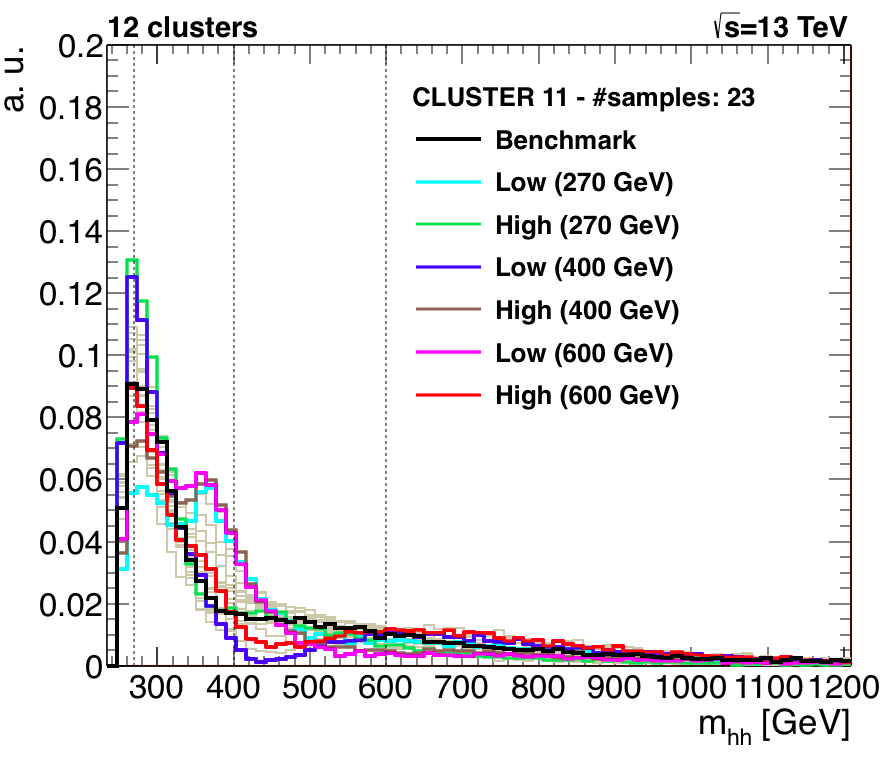}
\includegraphics[width=0.32\textwidth, angle =0 ]{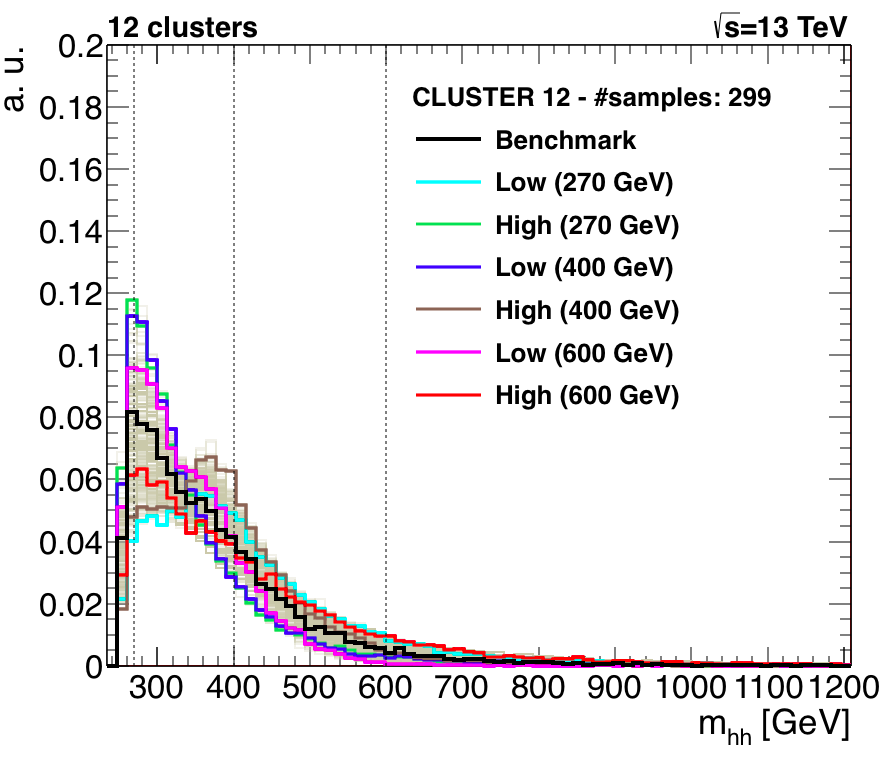}
\caption{\small The $\mHH$ distributions for the 12 clusters. The benchmark (in black color) and corresponding outliers (colored lines) are highlighted. The mass points $\mHH^1$, $\mHH^2$ and $\mHH^3$ are indicated by vertical dashed lines.
 \label{fig:app_out_mhh}}
\end{center}\end{figure}

The list of benchmarks is given in Table \ref{tab:bench}. We recommend the 12 benchmarks listed there to be the parameter space points targeted by experimental searches, in addition to the SM point. Fig.~\ref{fig:app_out_mhh} shows the $\mHH$ spectra for all the clusters together with the outliers, while Table~\ref{tab:out} provides the parameters of all the 72 outliers.

\begin{table}[h]
\centering
\small{
\begin{tabular}{r r r r r r r}
\toprule
Benchmark & $\kappa_{\lambda}$ & $\kappa_{t}$ & $c_{2}$	& $c_{g}$ & $c_{2g}$ \\
\midrule
1 &	7.5	 & 1.0	 &	-1.0	& 0.0	& 0.0 \\
2 &	1.0	 & 1.0	 &	0.5	& -0.8	& 0.6 \\
3 &	1.0	 & 1.0	 &	-1.5	& 0.0	& -0.8 \\
4 &	-3.5     & 1.5   &	-3.0	& 0.0	& 0.0 \\
5 &	1.0	 & 1.0	 &	0.0	& 0.8	& -1.0 \\
6 &	2.4	 & 1.0	 &	0.0	& 0.2	& -0.2 \\
7 &	5.0	 & 1.0	 &	0.0	& 0.2	& -0.2 \\
8 &	15.0     & 1.0	 &	0.0	& -1.0	& 1.0 \\
9 &	1.0	 & 1.0	 &	1.0	& -0.6	& 0.6 \\
10 &	10.0     & 1.5   &	-1.0	& 0.0	& 0.0 \\
11 &	2.4	 & 1.0	 &	0.0	& 1.0	& -1.0 \\
12 &	15.0     & 1.0	 &	1.0	& 0.0	& 0.0 \\ \midrule 
SM &	1.0      & 1.0	 &	0.0	& 0.0	& 0.0 \\
\bottomrule
\end{tabular}
}
\caption{\small Parameter values of the final benchmarks selected by the clustering procedure \cite{Dall'Osso:2015aia}. The third cluster is the one that contains the SM sample (defined by $\kappa_{\lambda}$ = $\kappa_{t}$ =1, $c_{2}$ = $c_{g}$ = $c_{2g}$ = 0). \label{tab:bench}}
\end{table}

Three of the clusters have benchmarks that do not obey the linear EFT  relation, however this does not present a problem for the interpretation of the results, assuming the latter. The phenomenological properties of parameter points within these clusters that belong to the linear realization of EWSB are still well approximated by the corresponding benchmarks.

Relevant properties of those three clusters are described below:
\begin{itemize}
\item In the scan we performed, Cluster 2 does not have representatives in the linear theory. The $\mHH$ extends above the TeV scale - therefore particular caution should be taken when interpreting the experimental results derived for the corresponding benchmark in the EFT (see Fig.~\ref{fig:app_out_mhh}). 

\item Cluster 3 includes the SM point and a large fraction of the points where only $\kappa_{\lambda}$ and $\kappa_t$ are modified, while the coefficients related to purely BSM operators are constrained to 0. 

\item Cluster 5 exhibits a doubly peaked structure in
$\mHH$, corresponding to maximal interference pattern and
associated with regions of minimal cross sections, which also
includes points of the linear case (see Fig.~\ref{fig:app_out_mhh}).

\end{itemize}

\begin{table}[h]
\centering
\footnotesize{
\begin{tabular}{|r r r r r|r r r r r|r r r r r|}
\toprule
 \multicolumn{5}{c}{Cluster 1} & \multicolumn{5}{c}{Cluster 2} & \multicolumn{5}{c}{Cluster 3}  \\[1mm] \hline
 $\kappa_{\lambda}$ & $\kappa_{t}$ & $c_2$ & $c_g$ & $c_{2g}$ & $\kappa_{\lambda}$& $\kappa_{t}$ & $c_2$ & $c_g$ & $c_{2g}$& $\kappa_{\lambda}$& $\kappa_{t}$ & $c_2$ & $c_g$ & $c_{2g}$ \\
 15.0  & 1.0 & -3.0 & 0.0 & 0.0   & 1.0 & 1.0 & 0.5 & 1.0 & 0.8   & -10.0 & 0.5 & 3.0 & 0.0 & 0.0 \\
 15.0  & 1.0 & -3.0 & 0.0 & 0$^{*}$ & 1.0 & 1.0 & 0.5 & 0.8 & 0.8 & 5.0 & 1.5 & -2.0 & 0.0 & 0.0 \\
 1.0   & 1.0 & 0.5  & 1.0 & 0.4   & 1.0 & 1.0 & 0.5 & -0.6 & 0.4  & 2.4 & 1.0 & -0.5 & 0.0 & 0.0 \\
 1.0   & 1.0 & 0.5  & 1.0 & 0.2   & 1.0 & 1.0 & 0.5 & -0.8 & 0.4  & 2.4 & 2.0 & 1.0 & 0.0 & 0.0 \\
 15.0  & 1.5 & -3.0 & 0.0 & 0.0   & 1.0 & 1.0 & 0.5 & -1.0 & 0.8  & 2.4 & 2.5 & 1.0 & 0.0 & 0.0 \\
 -10.0 & 0.5 & 1.0  & 0.0 & 0.0   & 1.0 & 1.0 & 0.5 & -1.0 & 1.0  & 1.0 & 2.5 & 1.5 & 0.0 & 0.0 \\\midrule
   \multicolumn{5}{c}{Cluster 4} & \multicolumn{5}{c}{Cluster 5} & \multicolumn{5}{c}{Cluster 6}  \\[1mm] \midrule
 $\kappa_{\lambda}$& $\kappa_{t}$ & $c_2$ & $c_g$ & $c_{2g}$ & $\kappa_{\lambda}$& $\kappa_{t}$ & $c_2$ & $c_g$ & $c_{2g}$& $\kappa_{\lambda}$& $\kappa_{t}$ & $c_2$ & $c_g$ & $c_{2g}$ \\
 5.0 & 2.0 & 3.0 & 0.0 & 0.0 & 5.0 & 1.0 & 0.0 & -0.6 & 0.6 & 10.0 & 2.5 & -2.0 & 0.0 & 0.0 \\
 1.0 & 1.5 & 3.0 & 0.0 & 0.0 & 1.0 & 1.0 & 0.5 & 1.0 & 0.6 & 5.0 & 1.0 & -0.5 & 0.0 & 0.0 \\
 1.0 & 1.0 & 2.0 & 0.0 & -0.4 & 1.0 & 1.0 & 0.0 & 0.2 & -0.8 & 7.5 & 1.5 & -1.0 & 0.0 & 0.0 \\
 1.0 & 1.0 & 0.0 & 0.4 & 0.6 & 1.0 & 1.0 & 0.5 & -1.0 & 0.2 & 2.4 & 1.0 & 0.0 & 0.4 & -0.4 \\
 1.0 & 1.0 & 0.0 & -0.2 & 0.4 & -2.4 & 1.5 & 2.0 & 0.0 & 0.0 & 5.0 & 2.5 & 1.0 & 0.0 & 0.0 \\
 -2.4 & 1.0 & 0.0 & -0.6 & 0.6 & -2.4 & 1.0 & 1.0 & 0.0 & 0.0 & 5.0 & 1.75 & 0.0 & 0.0 & 0.0 \\\midrule
  \multicolumn{5}{c}{Cluster 7} & \multicolumn{5}{c}{Cluster 8} & \multicolumn{5}{c}{Cluster 9}  \\[1mm] \midrule
   $\kappa_{\lambda}$& $\kappa_{t}$ & $c_2$ & $c_g$ & $c_{2g}$ & $\kappa_{\lambda}$& $\kappa_{t}$ & $c_2$ & $c_g$ & $c_{2g}$& $\kappa_{\lambda}$& $\kappa_{t}$ & $c_2$ & $c_g$ & $c_{2g}$ \\
 10.0 & 2.5 & 2.0 & 0.0 & 0.0 & 7.5 & 2.5 & -1.0 & 0.0 & 0.0 & 10.0 & 0.5 & -2.0 & 0.0 & 0.0 \\
 15.0 & 0.5 & 0.0 & 0.0 & 0.0 & 3.5 & 2.0 & 0.5 & 0.0 & 0.0 & 1.0 & 1.0 & 0.0 & -0.2 & -0.4 \\
 15.0 & 0.5 & 0.0 & 0.0 & 0$^*$ & 5.0 & 2.0 & 0.0 & 0.0 & 0.0 & 1.0 & 1.0 & 0.0 & -1.0 & -0.6 \\
 -12.5 & 0.5 & 0.5 & 0.0 & 0.0 & 7.5 & 2.0 & 3.0 & 0.0 & 0.0 & 1.0 & 1.0 & 0.5 & -1.0 & 0.0 \\
 5.0 & 1.0 & 0.0 & -0.2 & 0.2 & 1.0 & 1.0 & 0.5 & 0.4 & 0.0 & -5.0 & 1.0 & 2.0 & 0.0 & 0.0 \\
 7.5 & 1.75 & 0.0 & 0.0 & 0.0 & 1.0 & 1.0 & 1.5 & 0.4 & -0.4 & 1.0 & 1.0 & 0.5 & 0.4 & 0.2 \\\midrule
 \multicolumn{5}{c}{Cluster 10} & \multicolumn{5}{c}{Cluster 11} & \multicolumn{5}{c}{Cluster 12}   \\[1mm] \midrule
  $\kappa_{\lambda}$& $\kappa_{t}$ & $c_2$ & $c_g$ & $c_{2g}$ & $\kappa_{\lambda}$& $\kappa_{t}$ & $c_2$ & $c_g$ & $c_{2g}$& $\kappa_{\lambda}$& $\kappa_{t}$ & $c_2$ & $c_g$ & $c_{2g}$ \\
5.0& 1.0& 0.0 & -0.4& 0.4  & 5.0 & 2.0 & 1.0 & 0.0 & 0.0 & 3.5 & 0.75 & 0.5 & 0.0 & 0.0 \\
-7.5 & 0.5 & 0.5 & 0.0 & 0.0 & -2.4 & 2.5 & 3.0 & 0.0 & 0.0 & -5.0 & 1.0 & -1.0 & 0.0 & 0.0 \\
-12.5 & 1.5 & 3.0 & 0.0 & 0.0 & -3.5 & 2.0 & 2.0 & 0.0 & 0.0 & -2.4 & 2.5 & 1.0 & 0.0 & 0.0 \\
15.0 & 1.0 & -2.0 & 0.0 & 0.0 & -3.5 & 2.5 & 3.0 & 0.0 & 0.0 & 7.5 & 1.0 & 0.0 & 0.6 & -0.6 \\
-5.0 & 0.5 & 0.5 & 0.0 & 0.0 & -5 & 2.0 & 3.0 & 0.0 & 0.0 & 7.5 & 1.5 & 0.5 & 0.0 & 0.0 \\
-10.0 & 1.0 & 2.0 & 0.0 & 0.0 & -10.0 & 1.5 & 3.0 & 0.0 & 0.0 & -15.0 & 2.5 & 3.0 & 0.0 & 0.0 \\
\bottomrule
\end{tabular}
}
\caption{The parameter space coordinates of the outliers of the clusters whose
benchmarks are in Table~\ref{tab:bench}. 
\label{tab:out}}
\end{table}

\clearpage
\subsection{Experimental results from LHC Run I data taking period}

Recently both ATLAS and CMS collaborations performed searches for the 
non resonant production of Higgs boson pairs with 8\,TeV LHC data~\cite{Aad:2014yja, Aad:2015xja,  Aad:2015uka,  Khachatryan:2016sey}. 
The ATLAS collaboration considered only the SM-like kinematics for the signal.
The best upper limit of $\sigmaHH<0.69$\,pb is obtained in Ref.~\cite{Aad:2015xja} as a results of a combination of the four channels
$\HH \to \tau\tau\,b\bar{b}$, $\HH \to \gamma\gamma\,b\bar{b}$, $\HH \to b\bar{b}\,b\bar{b}$ and $\HH \to \gamma\gamma\,WW$. 
For the same hypothesis, the CMS collaboration provides a limit of $\sigmaHH<0.71$\,pb based on the results from the
$\HH \to \gamma\gamma\,b\bar{b}$ channel~\cite{Khachatryan:2016sey}. 

Reference~\cite{Khachatryan:2016sey} pushes further the exploration of the BSM parameter space for non-resonant $\HH$ production, by varying a subset of the parameter space, given by  $\kappa_{\lambda} ,\kappa_t $ and $c_2$. All in all, a grid of 124 points was generated. First a scan of $\kappa_{\lambda}$ was performed: $\kappa_{\lambda} = \pm$20, $\pm$15, $\pm$10, $\pm$5, 2.4, 1 (the SM point) and 0. In addition eight two-dimensional scans were done in the plane $(c_2,\kappa_t)$, for fixed values  of  $\kappa_{\lambda} = \pm$20, $\pm$15, $\pm$10, 1 and 0. In those, $c_2 = \pm$3, $\pm$2,  0 and  $\kappa_t = $ 0.75, 1, 1.25 were considered.  
One may notice that the scan in the $\kappa_t$ variable covers a smaller range than the scan that defined the clusters, but with a finer granularity.

The signals are searched for in two regions of $\mHH$, $<350$ and $\mHH >350$\,GeV, optimized to maximize the sensitivity for the SM-like search.
The observed limits on $\sigmaHH$ span between 1.36\,pb and 4.42\,pb depending on the point of the phase-space. By itself this result shows the 
importance of the kinematics of the BSM non-resonant production. 

The particularity of the $\HH \to \gamma\gamma\,b\bar{b}$ channel is to provide an excellent reconstruction of $\mHH$ with a resolution 
of $10-20$\,GeV and a rather constant signal efficiency for $250 < \mHH < 1000$\,GeV varying between 20\% and 30\%. 
Therefore this channel is well suited to resolve the details of the $\mHH$ spectrum and to challenge the cluster approach. This is what we discuss
in the following by applying the clustering technique to the analyzed parameter space.

\begin{figure}[hbt]\begin{center}
\includegraphics[width=1.0\textwidth, angle =0 ]{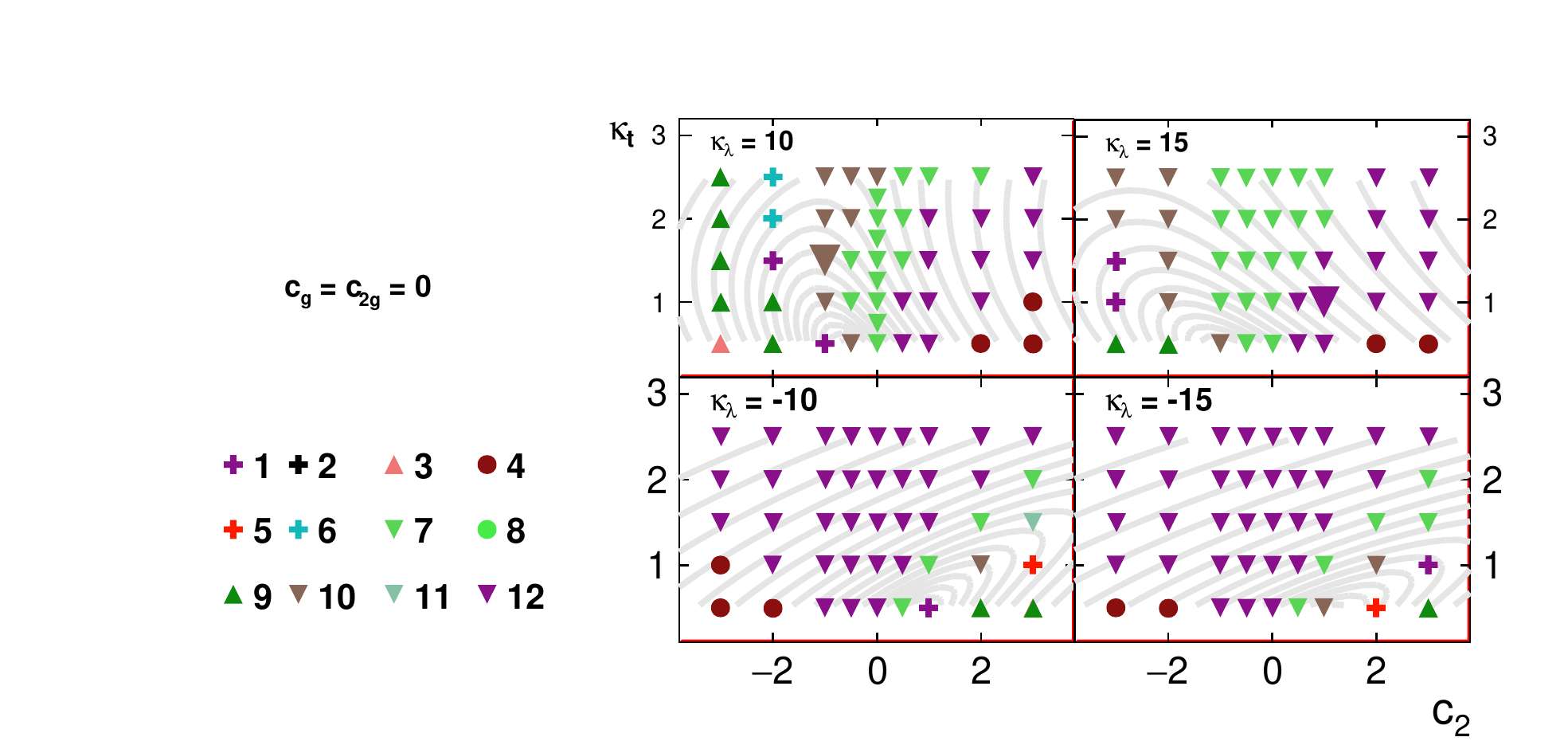}
\caption{\small Distribution of points in the $c_2 \times \kappa_t$ plane for different values of $\kappa_{\lambda}$ when $(c_g,\, c_{2g}) = (0,0)$.
Circles describe clusters whose benchmark has Higgs boson $p_T$ ($\pTH$) peaking around 100\,GeV. Downward-pointing triangles describe clusters where $\pTH$ is peaking around 50\,GeV or less, while upward-pointing triangles describe ones with $\pTH$ peaking around 150\,GeV or more. Finally, crosses describe clusters that show a double peaking structure in the $\pTH$ distribution. 
Larger markers indicate benchmark points. The gray lines correspond to iso-contours of constant cross section $\sigmaHH$.
See Fig.~8 in Ref.~\cite{Dall'Osso:2015aia} for more details. 
\label{fig:temp_aabb}}
\end{center}\end{figure}

From the 124 parameter space points for which we have experimental limits, only $N_{\rm samples } = 31$ coincide exactly with any of the parameter space points used to determine the clusters. However we note that other $N_{\rm samples} = 22$ extra points lie in between points that belong to the same cluster, while the gradient of cross section between them is smooth. This allows one to tag those intermediate points as belonging to the same cluster. 
As an example we show in Fig.~\ref{fig:temp_aabb} the distribution of the points used for the cluster definition in two-dimensional scans in 
 $(c_2,\kappa_t)$, for fixed values  of  $\kappa_{\lambda} = \pm$ 15 and $\pm$10. 
 Following the above defined algorithm the point $(\kappa_{\lambda} ,\kappa_t, c_2) = (15, 1.25, -2)$ is assigned as belonging to cluster 10. Similarly the point $(\kappa_{\lambda} ,\kappa_t, c_2) = (15, 0.75, -2)$  is not assigned to any cluster\footnote{Strictly speaking to assign those \textit{frontier} clusters we shall include them into the clustering procedure.}.

\begin{figure}[hbt]\begin{center}
\includegraphics[width=1.1\textwidth, angle =0 ]{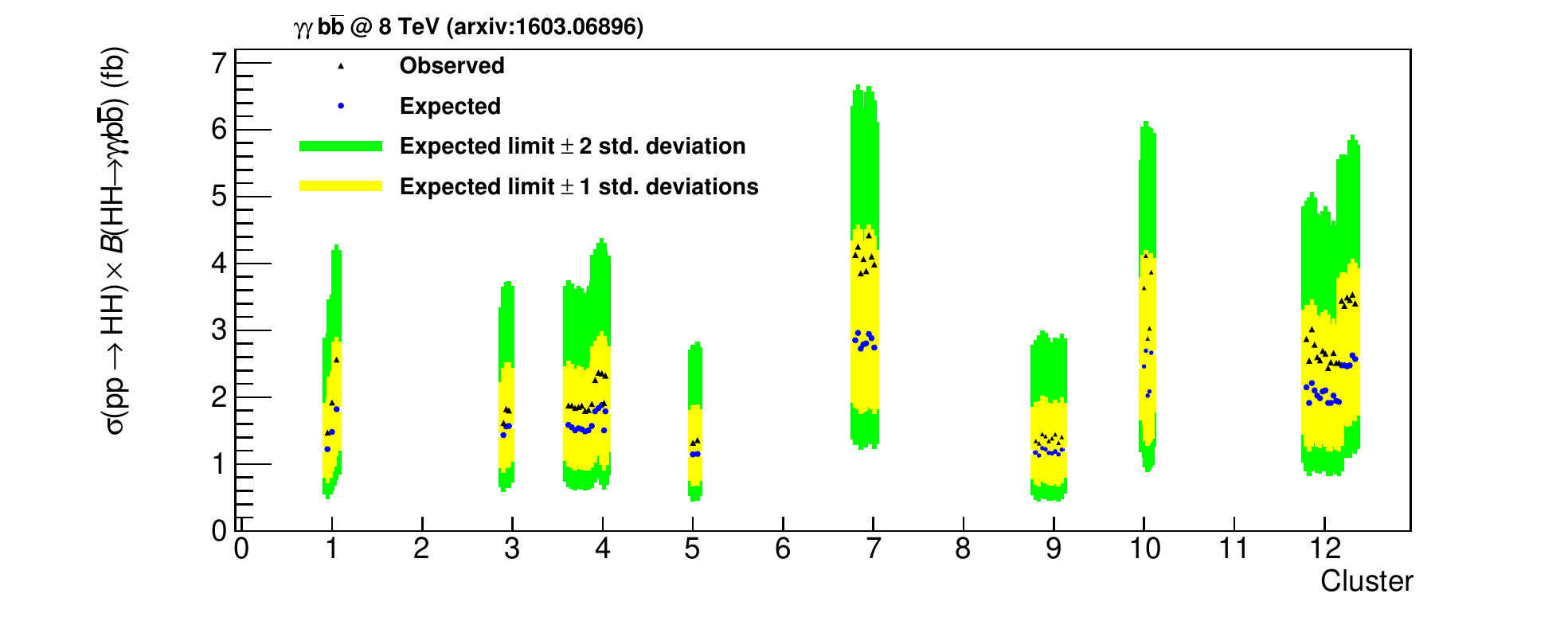}
\caption{\small Experimental limits from $\HH\to \gamma\gamma\,b\bar{b}$ search performed by the CMS collaboration~\cite{Khachatryan:2016sey}, organized by clusters following the procedure described in the text.  
The blue points correspond to the 95\% CL expected limits, while the green and yellow bands visualize one and two standard deviations around the latter.
\label{fig:lim_aabb}}
\end{center}\end{figure}

In Fig.~\ref{fig:lim_aabb} we show a total of 53 experimental limits at 95\% CL, organized by clusters. 
Unfortunately not all the clusters are equally populated and the statistics ranges between 0 and 19 samples. 
Still we have enough information to derive the first conclusions.

\begin{figure}[hbt]\begin{center}
\includegraphics[width=0.4\textwidth, angle =0 ]{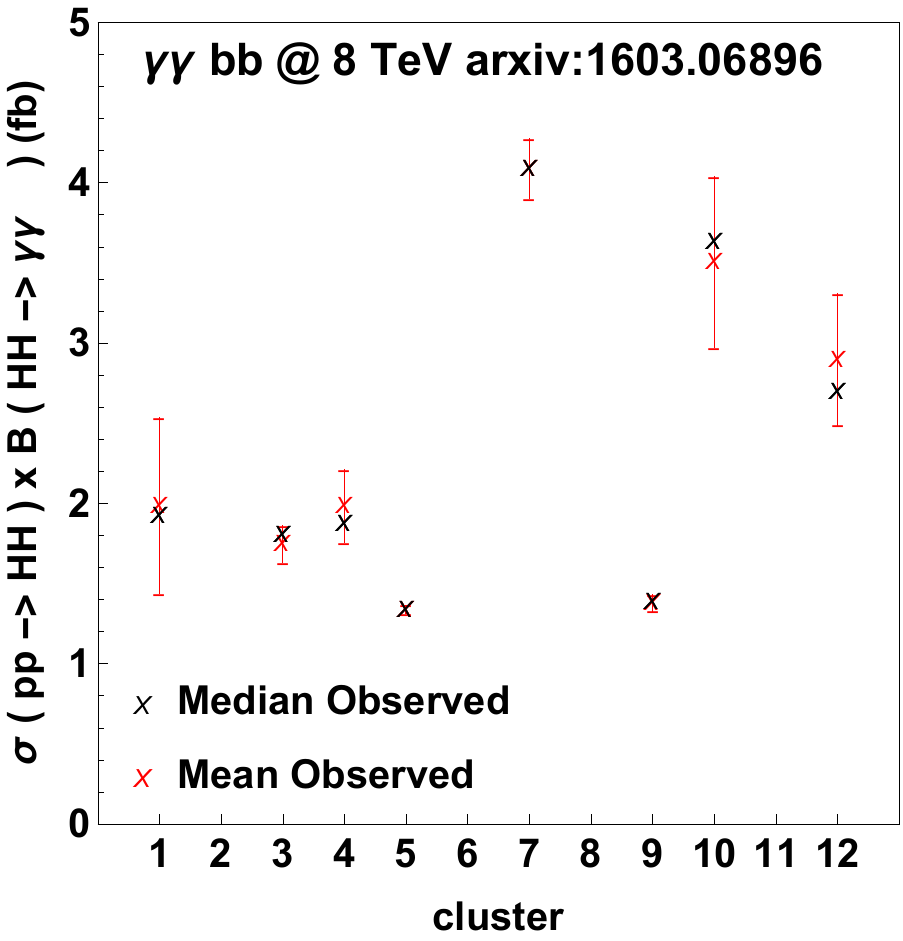}
\caption{\small Experimental limits from the $\HH \to \gamma\gamma\,b\bar{b}$ analysis performed by the CMS collaboration~\cite{Khachatryan:2016sey}, organized by clusters; see text for details.   \label{fig:mean_mass}}
\end{center}\end{figure}

Table~\ref{tab:mean} contains the mean values of the expected and observed experimental limits for each cluster and the standard deviations, defined with respect to the mean value. The same information is displayed in Fig.~\ref{fig:mean_mass}, for the observed limits only. 
As expected the largest limits correspond to threshold-like clusters, for instance cluster 7 and cluster 10.   
Clusters 4 and 12 seem to exhibit two sub-clusters each.  In the figure we also include the values of the medians. This estimator is less affected by the outliers than the mean. We observe that mean and median are close to each other in most cases.

\begin{figure}[hbt]\begin{center}
\includegraphics[width=0.42\textwidth, angle =0 ]{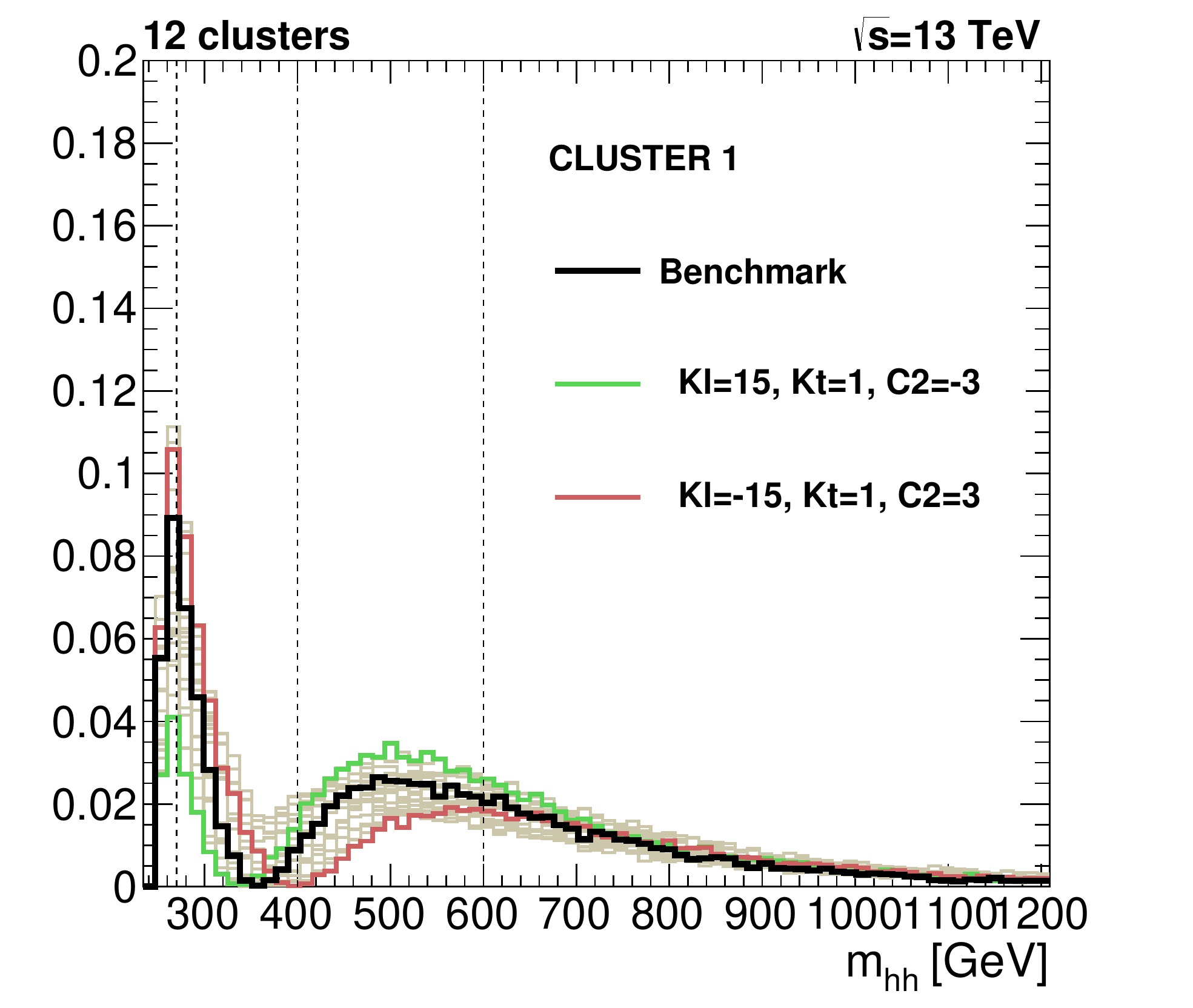}
\includegraphics[width=0.42\textwidth, angle =0 ]{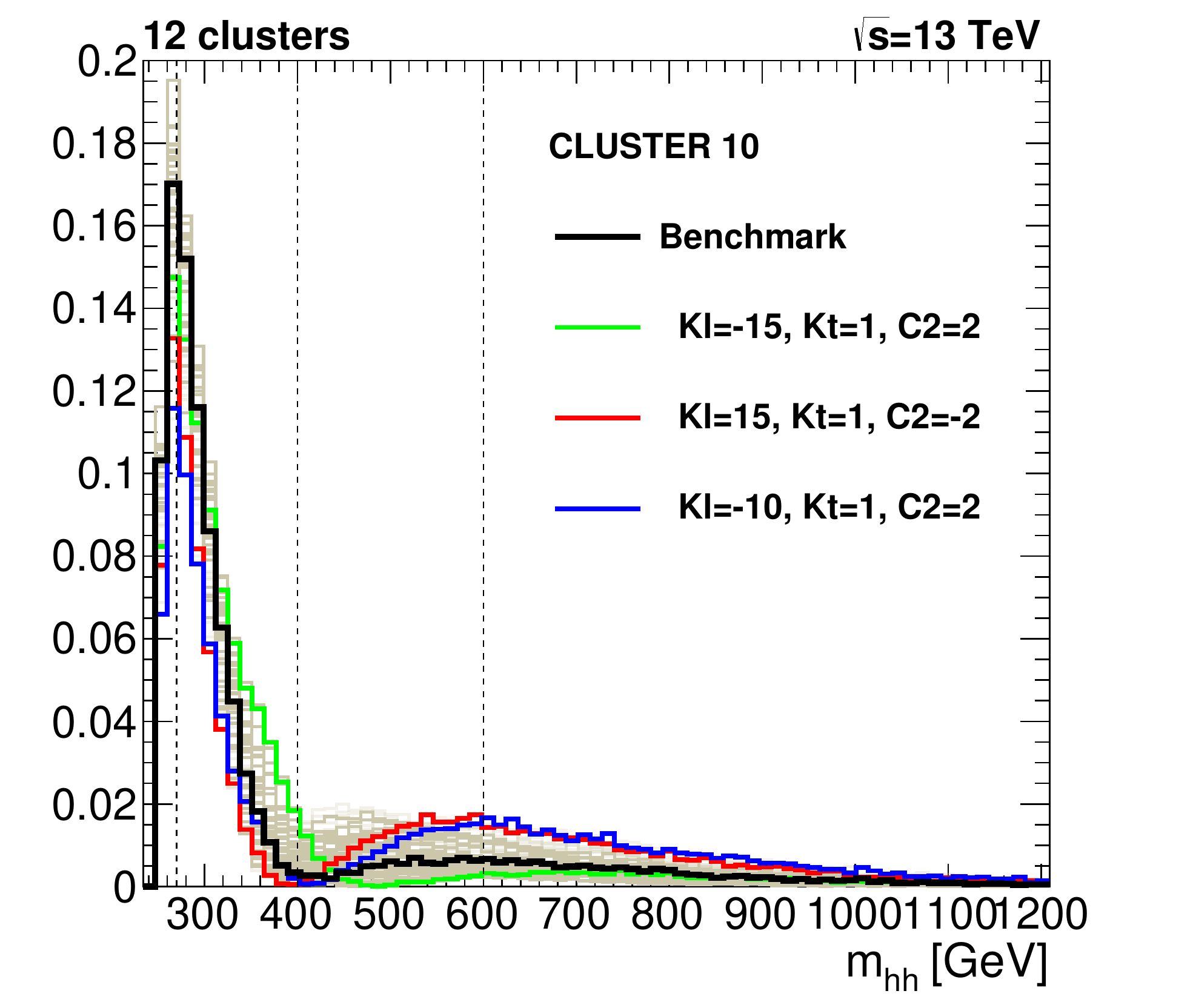}
\caption{\small The $\mHH$ distribution of the samples in cluster 1 and 10. The benchmark (black line) is compared with the 
points studied by~\cite{Khachatryan:2016sey} (bold color red and blue lines) that coincide with those considered in~\cite{Dall'Osso:2015aia}. \label{fig:out_mass_exp}}
\end{center}\end{figure}

As the observed limits are more susceptible to small data fluctuations their variance is bigger than the spread of the expected limits.  
The standard deviation derived for each cluster should be taken with care, as the parameter space scan in Ref.~\cite{Khachatryan:2016sey} was not done in a systematic way regarding the signal kinematic properties. 
For most of the clusters the relative size of the standard deviation does not exceed 5-10\%. For cluster 1 and 10 it increases up to 20\%. In Fig.~\ref{fig:out_mass_exp} we show the comparison of the $\mHH$ distributions. We observed as expected that the largest difference in limits corresponds to the largest difference in shapes. Therefore reweighting the benchmark to the outliers is an important element of an experimental analysis using this cluster technique.

By eye we see that the intra-cluster variance ($V_{\rm Intra, i}$) is smaller than the variance of limits between clusters (inter-cluster variance $V_{\rm Inter, i}$). 
To provide a more quantitative estimate of this phenomenon we use the Fisher-Snedecor test quite commonly used in biology for example to assess the compatibility of two medical tests. It is implemented via 

\begin{equation}
{\rm p}\mbox{-}{\rm value} = {\rm TMath::FDist}(V_{\rm Inter}/V_{\rm Intra, i}, N_{clus}, \#{\rm samples}). 
\end{equation}

It uses the ratio of inter-cluster variance and intra-cluster variance for each cluster. The number of degrees of freedom of the numerator is assumed to be $N_{clus} = 8$ and the denominator is the number of samples. The output of this function is a p-value on the hypothesis that both variances are statistically compatible. The results are provided in Table~\ref{tab:mean}. We observe that the clustering procedure is successful to reduce the variance of the phase-space and identify groups with similar experimental behavior in the complex phase-space of $\HH$ production final state.

\begin{table}[h]
\centering
\small{
\begin{tabular}{rcccccccc| c}
\toprule
Benchmark & 1  & 3 & 4 & 5  & 7  & 9 & 10  & 12 & All \\[1.5mm]
\midrule
\# samples  & 3 (1) & 3 (2) & 13 (4) & 2 (1) & 8 (5) & 9 (6) & 5 (2) & 19 (10) & 8 Means \\[1.5mm]
\midrule
Obs. Mean  (pb) & 1.92 & 1.75 & 1.87 & 1.34 & 4.08 & 1.39 & 3.63 & 2.69 & 2.33 \\[1.5mm]
$\sqrt{\mbox{Variance}}$ (pb) & 0.55 & 0.11 & 0.22 & 0.03 & 0.19 & 0.05 & 0.53 & 0.41 & 1.03 \\[1.5mm]
FS p-value & 6e-02 & 3e-05 & 6e-07 & 7e-07 & 5e-06 & 1e-12 & 4e-02 & 5e-04 & \\[1.5mm]
\midrule

Exp. Mean (pb) & 1.51 & 1.52 & 1.61 & 1.1484 & 2.84 & 1.18 & 2.39 & 2.18 & 1.80\\[1.5mm]
$\sqrt{\mbox{Variance}}$ (pb) & 0.30 & 0.08 & 0.15 & 0.005 & 0.09 & 0.04 & 0.32 & 0.25 & 0.61 \\[1.5mm]
FS p-value & 4e-02 & 8e-05 & 3e-06 & 4e-09 & 5e-07 & 3e-11 & 4e-02 & 6e-04 & \\[1.5mm]
\bottomrule
\end{tabular} 
}
\caption{\small Mean and standard deviations of the expected and observed limits for the 8 clusters analyzed in the $\HH\to \gamma\gamma\,b\bar{b}$ search performed by CMS collaboration. The label "\# samples" denotes the number of samples identified per cluster. The number in parentheses indicates how many of these were identified to be in the cluster from interpolation, as described in the text.  
The last column shows the total unweighted mean and the associated standard deviation between the means of the 8 clusters. The Fisher-Snedecor p-values are also provided.
\label{tab:mean}}
\end{table}

\clearpage
\section{Analytical parametrization of the cross section}
\label{sec:cx}

The effect of anomalous couplings on the di-Higgs production cross section can be written in the form of a ratio $\RHH$ between the cross section of the BSM model and the SM cross section using a vector of numerical coefficients  $\vec{A}=[A_1, ... , A_{15}]$:  

\begin{equation}
\begin{split}
\RHH \equiv \frac{\sigmaHH}{\sigmaHH^{\rm SM}} \overset{LO}{=}   & A_1\, \kappa_t ^4 + A_2\, c_2^2 + (A_3\, \kappa_t^2 + A_4\, c_g^2)\, \kappa_{\lambda}^2  + A_5\, c_{2g}^2  \\
 & + ( A_6\, c_2 + A_7\, \kappa_t \kappa_{\lambda} )\kappa_t^2  + (A_8\, \kappa_t \kappa_{\lambda} + A_9\, c_g \kappa_{\lambda} ) c_{2}  \\
 & + A_{10}\, c_2 c_{2g}  + (A_{11}\, c_g \kappa_{\lambda} + A_{12}\, c_{2g})\, \kappa_t^2  \\
 & + (A_{13}\, \kappa_{\lambda} c_g + A_{14}\, c_{2g} )\, \kappa_t \kappa_{\lambda} + A_{15}\, c_{g} c_{2g} \kappa_{\lambda} \,.
\end{split}
\label{eq:cx}
\end{equation} 

To obtain the total cross section for each point in the parameter space described by the couplings of the Lagrangian 
(\ref{eq:lag}), we recommend to use the relation:

\begin{equation}
\sigmaHH = \sigmaHHSMNNLO \cdot \RHH\,,
\label{eq:cxNNLO}
\end{equation} 

\noindent where $\sigmaHHSMNNLO$ is the state-of-the-art SM cross section, calculated including QCD corrections at Next-to-Next-to-Leading Order (NNLO) and matched to Next-to-Next-to-Leading Log resummations (NNLL),  this result can be found on~\cite{MelladoGarcia:2150771}.  It was obtained by several independent calculations~\cite{Dawson:1998py,Borowka:2016ehy, Liu-Sheng:2014gxa,Grigo:2014jma,Grigo:2015dia,deFlorian:2015moa,deFlorian:2013jea}.

\subsection{Definition of the procedure to extract the coefficients}\label{sec::procedure}
The coefficients $\vec{A}$ can be extracted from a fit to the cross sections estimated by MC integration in different points of the BSM parameter space. The choice of those input points is critical in order to obtain an accurate parametrization. One needs to assure a sufficient sensitivity of the $\HH$ cross section to all coefficients in Eq.~(\ref{eq:cx}) and at the same time scan a broad enough range such as to obtain a control on the error in the fit and its internal consistency. To this end, we explore the various directions in the parameter space by studying two-dimensional (2D) planes, starting with the SM-like plane $(\kappa_{\lambda} ,\kappa_t )$ and the $(c_2,\kappa_t)$ plane. We follow the relation between the Higgs-gluon contact interactions that comes from the 
linear dimension-6 EFT formalism ($c_g = - c_{2g}$) to select two other 2D planes $(c_2, c_g)$ and $(\kappa_{\lambda} , c_g)$. 
Finally, the ambiguities from the EFT relation are removed by scanning the planes $(c_g, c_{2g})$ and $(c_{2g}, c_{2})$.  
Parameters that are not mentioned here are set to their SM values, such that the SM benchmark point is present in all the two-dimensional subsets of the point set. 
The final set is composed of 251, 266, 265, 261 and 169 points, corresponding to LHC center-of-mass (CM) energies of 7, 8, 13, 14, and 100 TeV, respectively. We verified that for each CM energy the number of samples is always sufficient to constrain the coefficients with high precision, as shown below.

The components of $\vec{A}$ are extracted by maximizing the likelihood simultaneously for all the coefficients, taking into account all the points, \emph{i.e.}, minimizing
\begin{equation}
\log L(\vec{A}) = \sum_{i\, \subset\, \mbox{event set} } \left(  \frac{\sigmaHHMC^i-\sigmaHH^i(\vec{A})}{\delta \sigmaHHMC^i} \right)^2\,,
\label{eq:lik}
\end{equation}

\noindent
where $i$ runs over all the points in the set, $\sigmaHHMC^i$ is the corresponding cross section calculated by the simulation, $\sigmaHH^i(\vec{A})$ is the cross section in the same parameter space point calculated via Eq.~\ref{eq:cx} and $\delta \sigmaHHMC^i$ is the statistical MC uncertainty assigned to each point. The minimization of $\log L(\vec{A})$ is performed with MINUIT and the results are cross-checked with a fit performed with {\tt Mathematica}. 

The cross sections are calculated using the Madgraph\_aMC@NLO model also used for the signal shapes. We generate 10,000 events per point $i$.
As proton setting we employ the central PDF of the PDF4LHC15\_nlo\_mc\_pdfas~\cite{Carrazza:2015hva, Butterworth:2015oua, Dulat:2015mca,Harland-Lang:2014zoa, Ball:2014uwa} set\footnote{The settings used for the calculation of the cross section follow the recommendations of Ref.~\cite{YR4}. 
One may observe that the settings used earlier to cluster the shapes were slightly different (see Ref.~\cite{Dall'Osso:2015aia} and section \ref{sec:clus}). This difference has no impact on the discussion since the clustering procedure is not sensitive to small changes in QCD parameters.}, 
the strong coupling is taken as $\alpha_s(m_Z) = 0.118$, and the factorization and renormalization scales are fixed to $\mHH/2$. The input masses are $\mH$ = 125\,GeV, $m_t$ = 173.18\,GeV, and $m_b$ = 4.75\,GeV. 
Each point is simulated with a different random seed to avoid statistical correlations between the points. 
 
The size of $\delta \sigmaHHMC^i/ \sigmaHHMC^i $ is estimated a posteriori after a first fit by looking on the pulls between the simulated cross sections and the interpolated ones, requiring ${\rm RMS}\approx 1$ for the residuals of the fit\footnote{This procedure is well justified, since it can be shown that the parametrisation of Eq. (\ref{eq:cx}) is exactly correct at LO. It is just the integration error that leads to a non-vanishing difference between the fit and the MC cross section.}. With this procedures we find $\delta \sigmaHHMC^i/ \sigmaHHMC^i = 0.03\%$ is a good estimate of the uncertainties at all CM energies.
The SM point is a particular case: by definition $\RHH(SM) = 1$ with no uncertainty. In practice to avoid infinite values in the likelihood we define $\delta \sigmaHHMC^{i = SM} / \sigmaHHMC^{i = SM}  \equiv 0.01\%$. 


\subsection{Fit results}

\begin{figure*}[hbt]\begin{center}
\includegraphics[width=0.32\textwidth, angle =0 ]{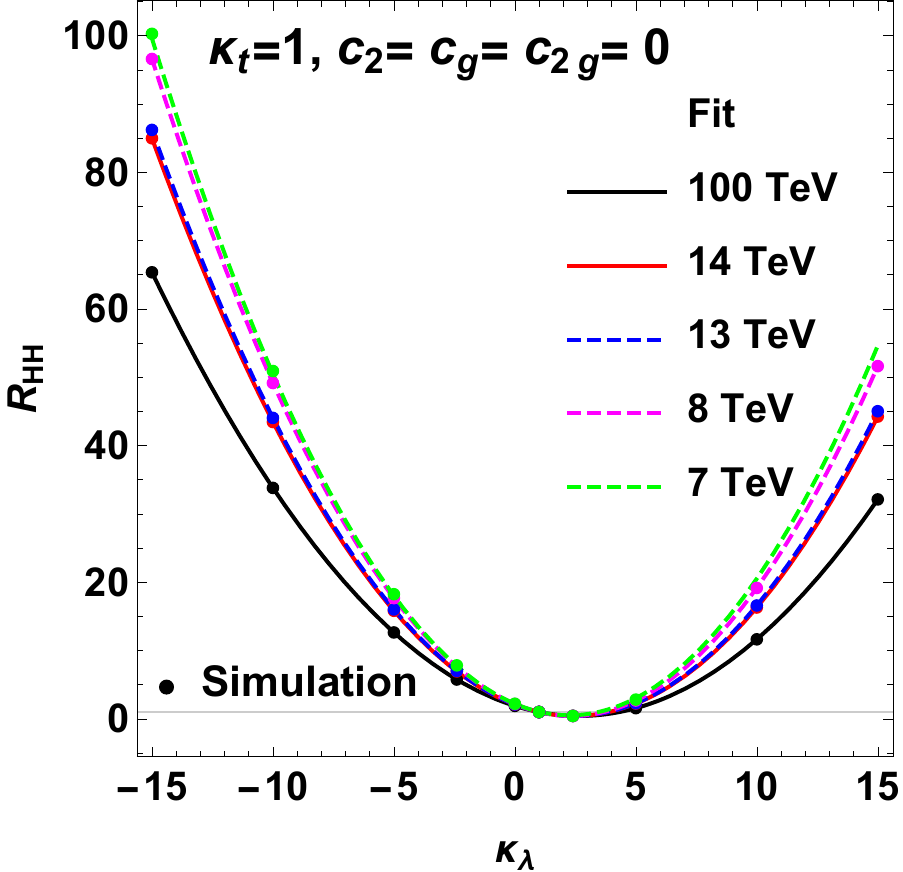}
\includegraphics[width=0.32\textwidth, angle =0 ]{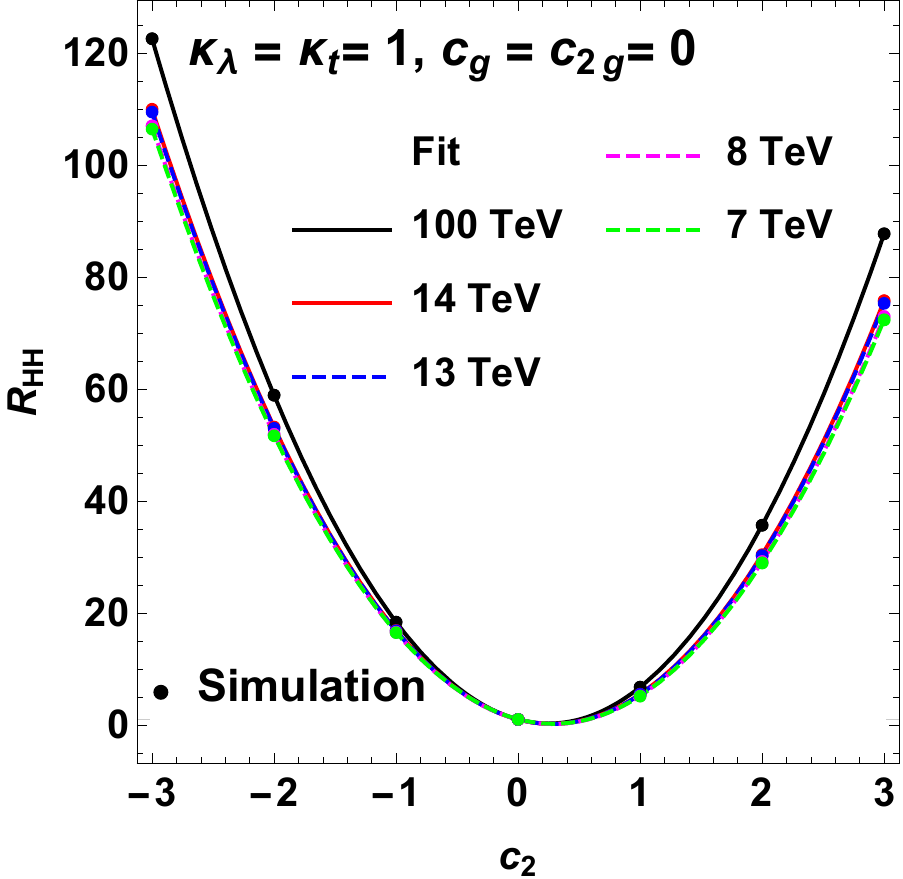}
\includegraphics[width=0.31\textwidth, angle =0 ]{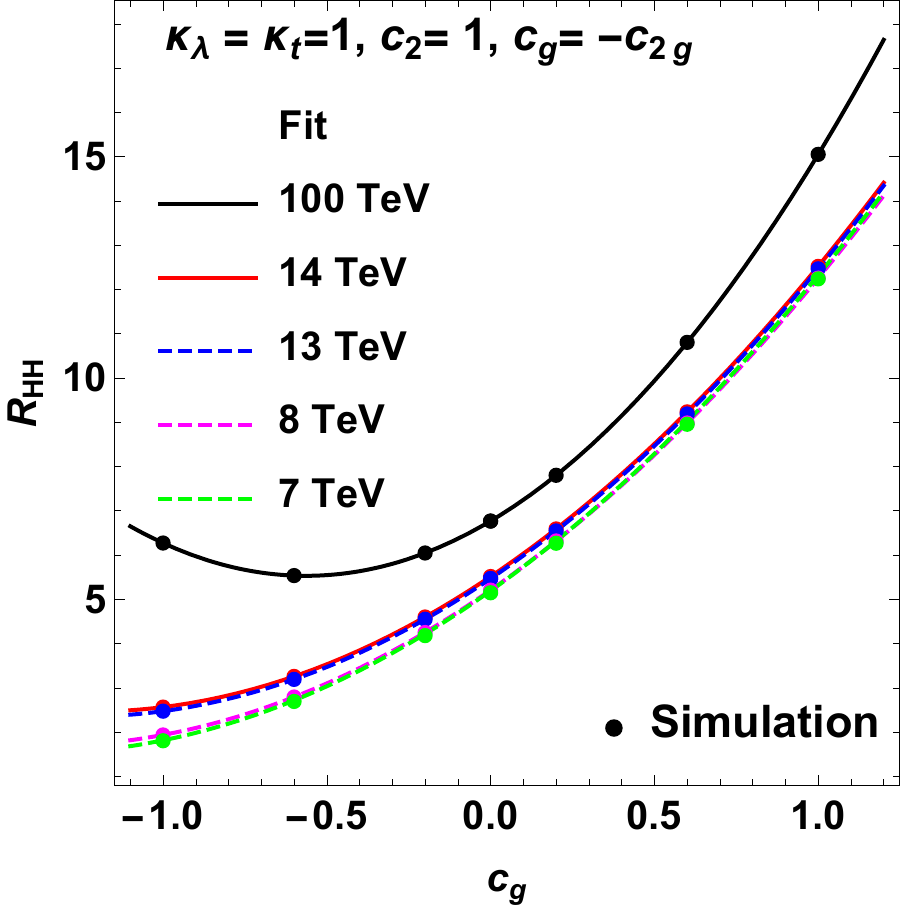}
\includegraphics[width=0.3\textwidth, angle =0 ]{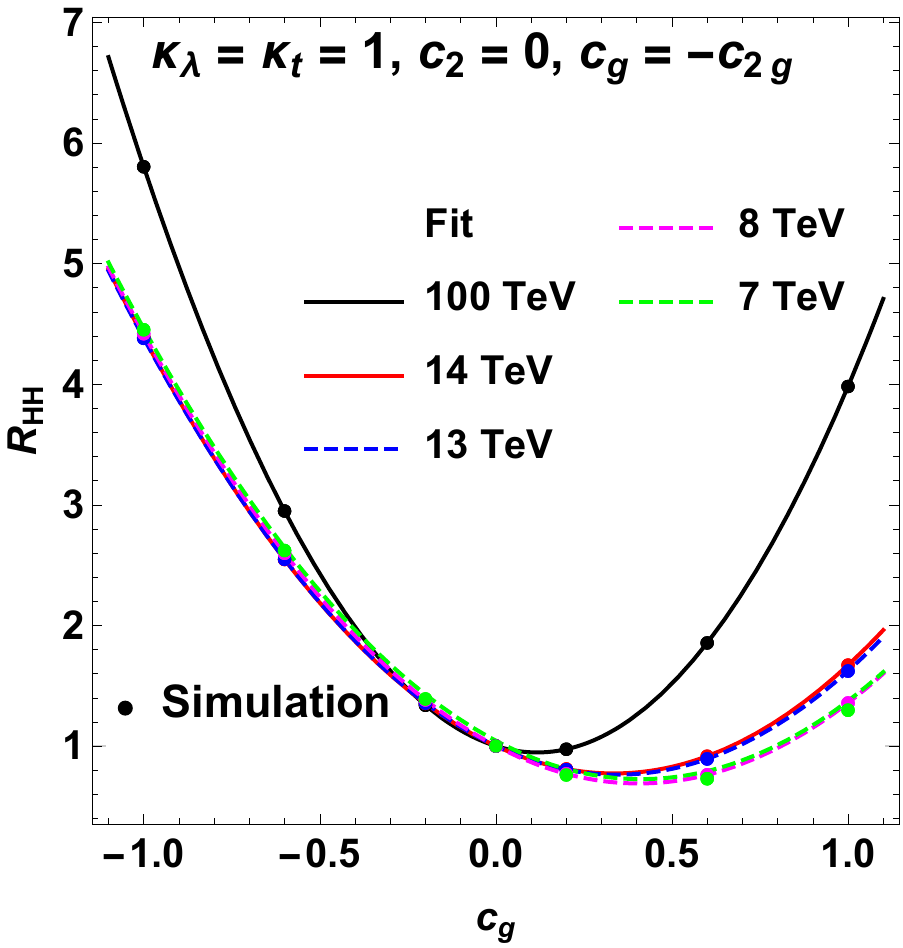}
\includegraphics[width=0.32\textwidth, angle =0 ]{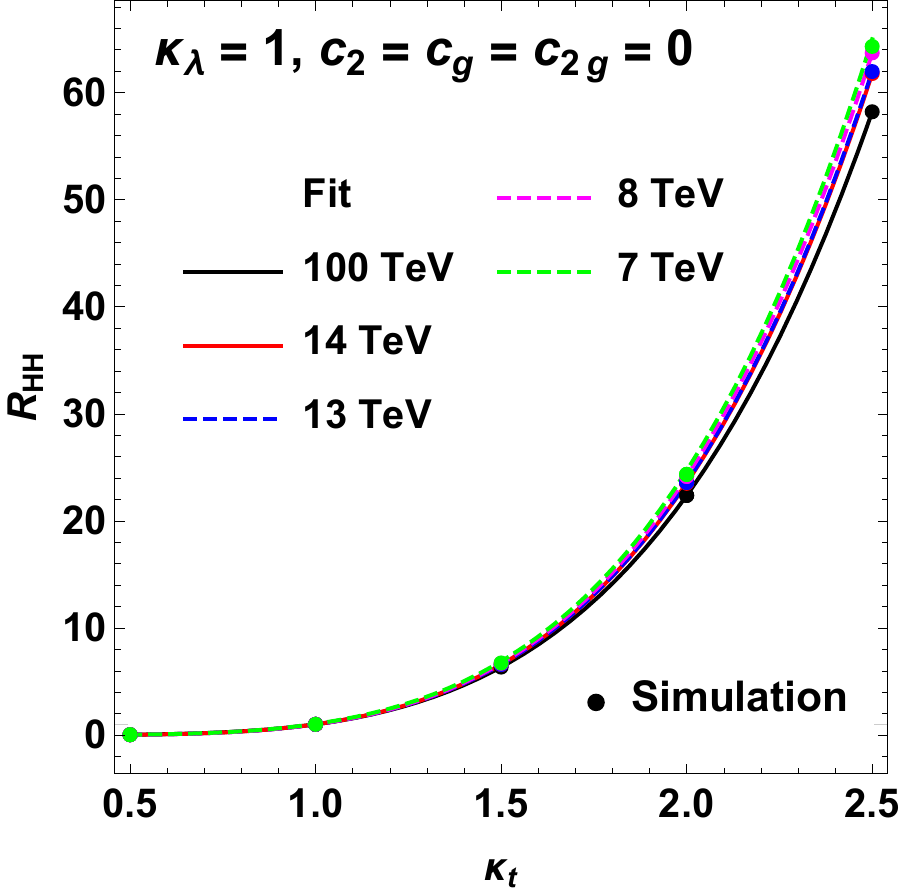}
\includegraphics[width=0.32\textwidth, angle =0 ]{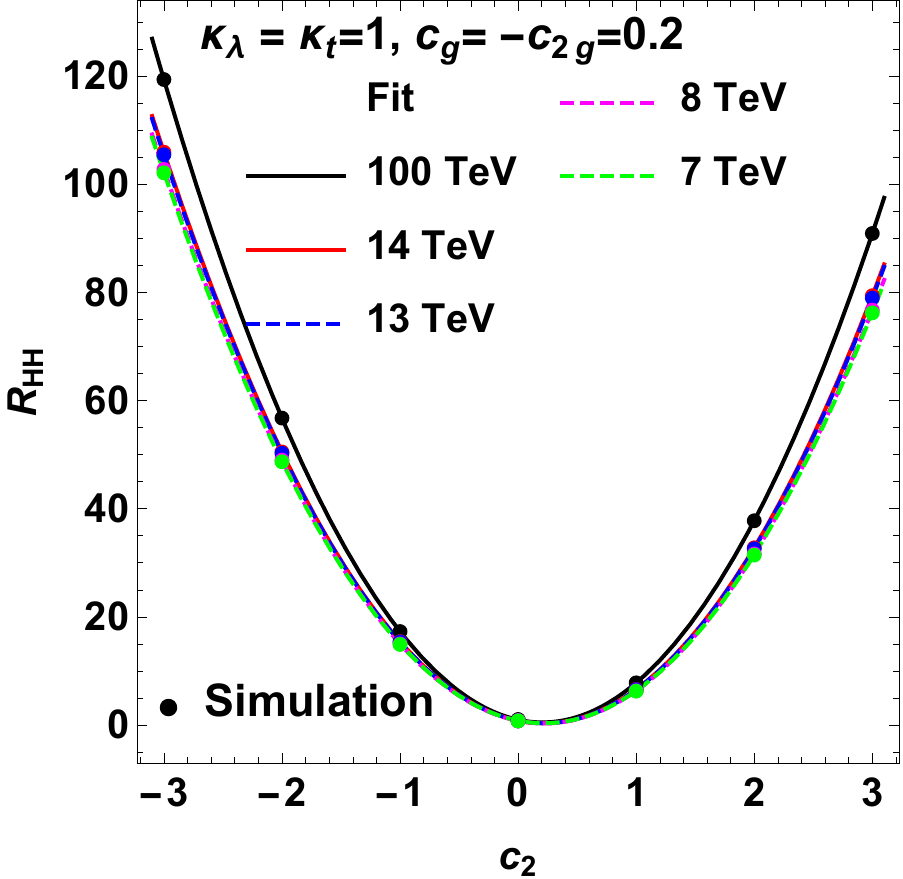}
\caption{\small Comparison of the cross sections predicted by Eq. \ref{eq:cx} with the MC cross sections for different combinations of parameters.
 \label{fig:comp}}
\end{center}\end{figure*}

The central values for $\vec{A}$ are shown in Table \ref{tab:coef}.
It is not trivial to identify a general trend in the behavior of the coefficients as a function of the CM energy.
We may still observe that the coefficients $A_1$, $A_3$, and $A_7$ related to the SM like Feynman diagrams ((a) and (b) in Fig.~\ref{fig:dia}) and their interference term decrease in magnitude with CM. The coefficients related to the pure BSM diagrams $A_2$ and $A_5$ ((c) and (e) in Fig. \ref{fig:dia}) in contrary increase, which can be understood from the fact that they correspond to genuinely higher dimensional contributions. The coefficient $A_4$ ((d) in Fig. \ref{fig:dia}) mixing a BSM operator and SM-like one is rather stable. The trend of the other coefficients corresponding to interference terms are more complex to describe.
Fig. \ref{fig:comp} shows the comparison of the MC cross section with the result of the cross section Formula~\ref{eq:cx}, using the coefficients of Table \ref{tab:coef}. We display $\RHH$ as function of different couplings in sub-spaces of the six planes used to fix the latter formula for the LHC at 13\,TeV. 
The order of magnitude and general behavior of the minima of $\RHH$ is very similar for the LHC running at 7-14\,TeV. Differences may be observed when considering a large jump from 14 to 100\,TeV energy in CM.

\begin{figure*}[hbt]\begin{center}
\includegraphics[width=0.7\textwidth, angle =0 ]{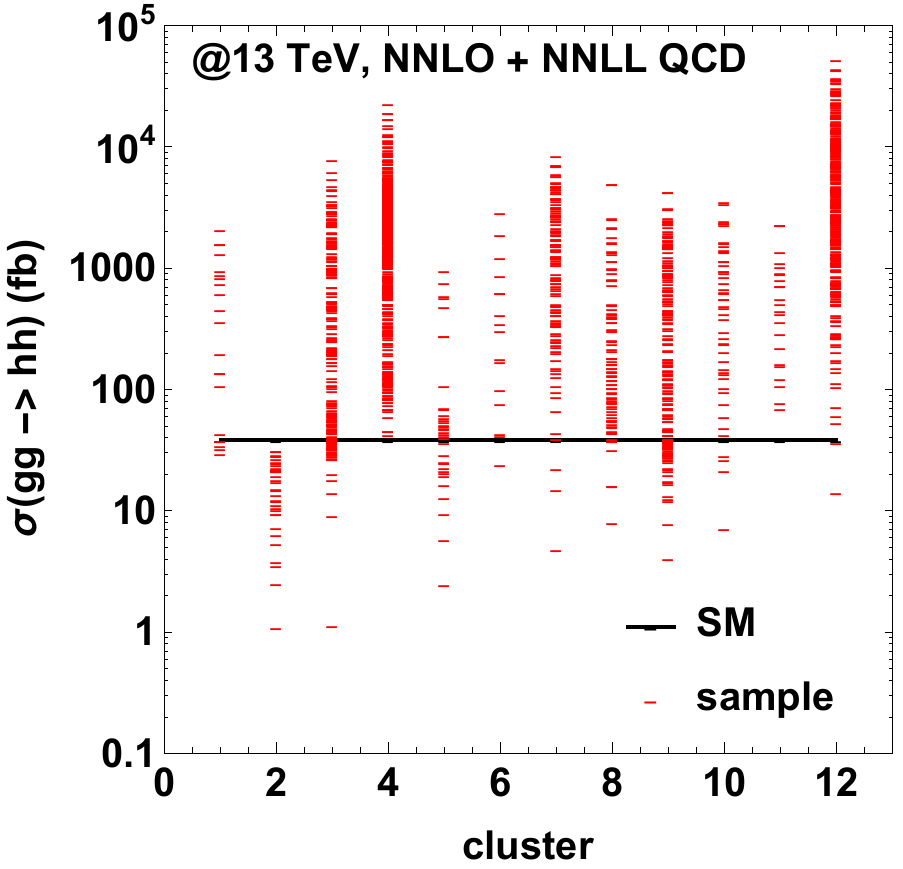}
\caption{\small 
Cross sections for the parameter-space points, depicted by red markers and grouped into 12 clusters. The black horizontal line shows the SM cross section.
\label{fig:XS}
}
\end{center}\end{figure*}

Finally, we illustrate in Fig. \ref{fig:XS} that, although kinematics and total cross section are correlated on one hand, the same topology can be obtained for points with cross sections that differ by orders of magnitude but, on the other hand, points with the same total cross section can feature very different kinematics.

\begin{table}[h]
\centering
\footnotesize{
\begin{tabular}{r r r r r r}
\toprule
$\sqrt{s} $ & 7\,TeV & 8\,TeV & 13\,TeV & 14\,TeV & 100\,TeV\\
\midrule
$A_1$  &  2.21  &  2.18  &  2.09  &  2.08  &  1.90  \\
$A_2$  &  9.82  &  9.88  &  10.15  &  10.20  &  11.57  \\
$A_3$  &  0.33  &  0.32  &  0.28  &  0.28  &  0.21  \\
$A_4$  &  0.12  &  0.12  &  0.10  &  0.10  &  0.07  \\
$A_5$  &  1.14  &  1.17  &  1.33  &  1.37  &  3.28  \\
$A_6$  &  -8.77  &  -8.70  &  -8.51  &  -8.49  &  -8.23  \\
$A_7$  &  -1.54  &  -1.50  &  -1.37  &  -1.36  &  -1.11  \\
$A_8$  &  3.09  &  3.02  &  2.83  &  2.80  &  2.43  \\
$A_9$  &  1.65  &  1.60  &  1.46  &  1.44  &  3.65  \\
$A_{10}$  &  -5.15  &  -5.09  &  -4.92  &  -4.90  &  -1.65  \\
$A_{11}$  &  -0.79  &  -0.76  &  -0.68  &  -0.66  &  -0.50  \\
$A_{12}$  &  2.13  &  2.06  &  1.86  &  1.84  &  1.30  \\
$A_{13}$  &  0.39  &  0.37  &  0.32  &  0.32  &  0.23  \\
$A_{14}$  &  -0.95  &  -0.92  &  -0.84  &  -0.83  &  -0.66  \\
$A_{15}$  &  -0.62  &  -0.60  &  -0.57  &  -0.56  &  -0.53  \\
\bottomrule
\end{tabular}
}
\caption{\small 
Central values for the coefficients entering $R_{hh}$ (Eq. 
(\ref{eq:cx})), employing the anomalous couplings parametrization (Eq.~\ref{eq:lag}).
 \label{tab:coef}}
\end{table}

\clearpage

\subsection{Uncertainties}

The different sources of uncertainties considered in this analysis are: statistical uncertainties on the MC samples, uncertainty in QCD parameters (proton PDF and $\alpha_s$) as well as missing order uncertainties.

The statistical uncertainties in the cross section for each sample $i$ predicted by MC integration was estimated in Section \ref{sec::procedure} to be 0.03\%. The resulting impact on the coefficients $\vec{A}$ was observed to be negligible.

The cross section uncertainties due to the different parton distribution functions (PDFs) and $\alpha_{s}$
are obtained following the recommendation for Run 2 provided by Ref. \cite{Butterworth:2015oua}.
We use the MC PDF set PDF4LHC15\_nlo\_mc\_pdfas with $N_{\rm rep} =$ 100 replicas and $\alpha_s(m_Z) = 0.118$. Moreover, we consider two extra replicas 
 with $\alpha_s(m_Z) = 0.1165$ and $\alpha_s(m_Z) = 0.1195$.

To estimate the PDF uncertainty in $\RHH$ we calculate for a sample $i$ and a replica $j$ the deviation

\begin{equation}
\delta_{\rm PDF}^{i,j} \equiv 1- \frac{\RHH^{i,j}}{\RHH^{i, CV}}
\end{equation}
with respect to the value of the ratio calculated with the central value of PDF4LHC15\_nlo\_mc\_pdfas, $\RHH^{i, CV}$. 
The PDF uncertainty for $i$ is then obtained as

\begin{equation}
\delta_{PDF}^{i} \equiv \sqrt{ \frac{1}{N_{rep}-1}  \sum_{j=1}^{N_{rep}} (\delta_{PDF}^{i,j})^2}.
\end{equation}

The $\delta_{\alpha_S}^i$ uncertainty is estimated as the relative difference between two replicas obtained with modified values of $\alpha_s(m_{\rm Z})$: 0.1165 and 0.1195.
The total uncertainty on $\RHH^i$ can finally be obtained as

\begin{equation}
\delta \RHH^i \equiv \sqrt {(\delta_{\rm PDF}^{i})^2 +(\delta_{\alpha_S}^i)^2}.
\end{equation}

The uncertainty due to the QCD parameters related to proton settings in the total cross section is a function of the signal topology. To good approximation, all samples within a cluster probe the same topology. We present in Table~\ref{tab:error} the impact of the uncertainties on the 12 benchmarks of Table~\ref{tab:bench} for different CM energies. 
We observe that the QCD uncertainties cancel out in the ratio down to a residual few per mill. In consequence, the uncertainty in $\RHH$ due to limited MC statistics and QCD parameters are negligible to very good approximation. 

\begin{table}[h]
\centering
\small{
\begin{tabular}{rcccccccccccc}
\toprule
Benchmark & 1 & 2 & 3 & 4 & 5 & 6 & 7 & 8 & 9 & 10 & 11 & 12 \\[1mm]
\midrule
$\sqrt{s}$ &  \multicolumn{12}{c}{$\delta \RHH^i$ (\%)}\\[1mm]
\midrule
8\,TeV & 0.1 & 0.2 & 0.0 & 0.1 & 0.1 & 0.1 & 0.2 & 0.1 & 0.1 & 0.2 & 0.0 & 0.1 \\[1mm]
13\,TeV & 0.1 & 0.3 & 0.0 & 0.0 & 0.1 & 0.1 & 0.2 & 0.1 & 0.1 & 0.2 &  0.0  & 0.1 \\[1mm]
14\,TeV & 0.1 & 0.3 & 0.1 & 0.0 & 0.2 & 0.1 & 0.2 & 0.2 & 0.1 & 0.2 & 0.0 & 0.1 \\[1mm]
100\,TeV & 0.3  & 1 & 0.2 & 0.0 & 0.8 & 0.2 & 0.4 & 0.5 & 0.5 & 0.2 & 0.1 & 0.3 \\
\bottomrule
\end{tabular} 
}
\caption{\small Total theory uncertainty on the ratio $\RHH^i$ (in \%), including PDF and $\alpha_S$ variations, for each of the benchmark points $i$ of table~\ref{sec:bench} and the four center of mass energies we consider. The null entries correspond to points where $\delta \RHH^i$  is smaller than 0.05\%.  
\label{tab:error}}
\end{table}

Finally, we tried to estimate the impact of the missing orders on $\RHH$ calculated at LO. The $K$-factor for the total cross section is found to be fairly flat in the five parameter space directions we scan here when calculated at NLO QCD~\cite{Grober:2015cwa}. Consequently it almost cancels out in $\RHH$. The largest observed variation of 5\% in the infinite top mass limit  appears for the extreme BSM case of a sizable contact interaction among two Higgs bosons and gluons. 

This modest value suggests that the theory uncertainties in the total cross section
that are due to missing orders can be well approximated by the theory uncertainties assumed for the cross
section normalization $\sigmaHHSMNNLO$, as recommended in Ref.~\cite{MelladoGarcia:2150771}.

\begin{figure}[htb]\begin{center}
\includegraphics[width=0.50\textwidth, angle =0 ]{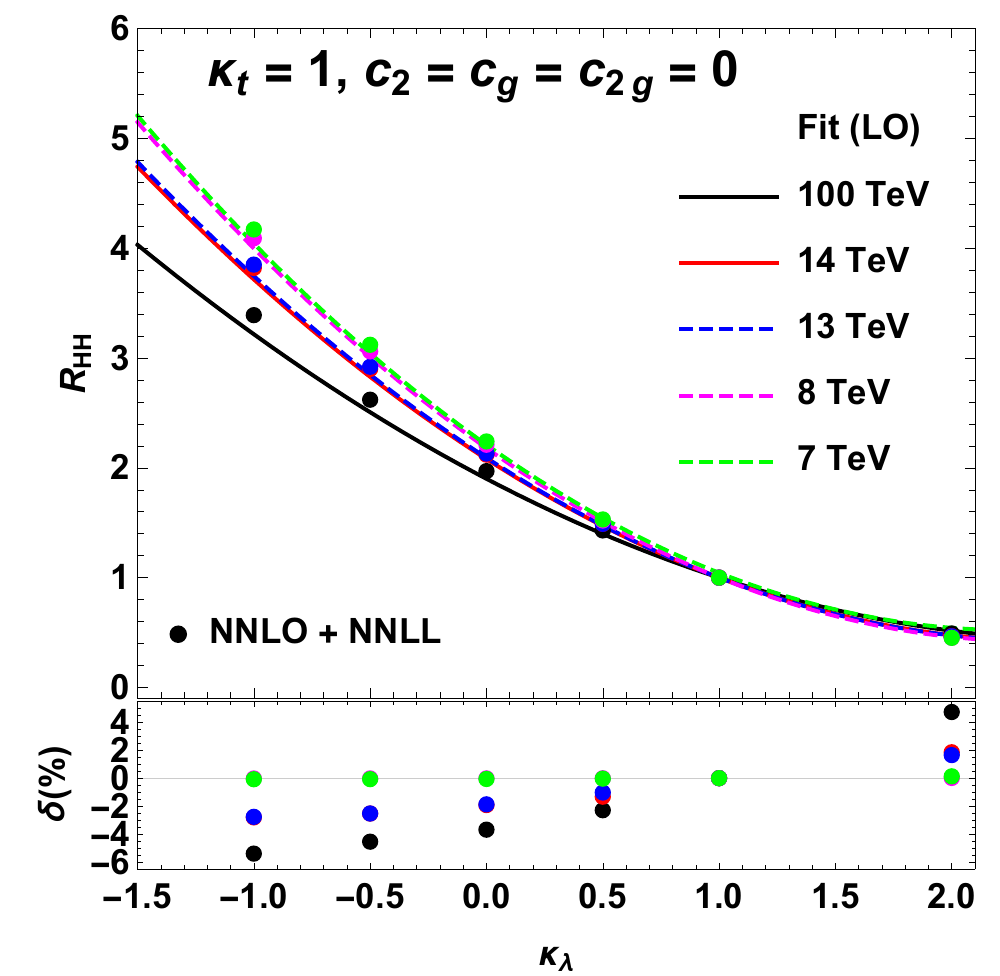}
\caption{\small Comparison between $\RHH$ obtained from Eq.~\ref{eq:cxNNLO} (line) and the ones obtained with NNLO+NNLL accuracy (points) ~\cite{deFlorian:2015moa}. 
 \label{fig:error}}
\end{center}\end{figure}

We also compared our predictions at LO in a narrow range of variations in the trilinear self-coupling $(\kappa_{\lambda} \in [-1,2])$, to the $\RHH$ calculated including QCD corrections up to NNLO and NNLL, which corresponds to the state of the art for the SM calculation~\cite{MelladoGarcia:2150771,deFlorian:2015moa,deFlorian:2013jea}. We find the maximum deviation from our predictions to be 6\% at a CM energy of 100\,TeV at the boundaries of the inspected range, see Fig.~\ref{fig:error}. 

We observe therefore a maximal impact of the order of 5\% from missing orders in QCD on $\RHH$ calculated at LO. Still, it is  hard to use this observation to derive a precise numerical recommendation on the (modest) uncertainty to be used for each point of the parameter space. Indeed the $K$-factor was obtained in the infinite top mass approximation that is challenged for large values of $\mHH$. Moreover a non-negligible variation was observed within the BSM parameter space. 
In any case, since the missing order uncertainties (including $m_t$ effects) on $\sigmaHHSMNNLO$ appear to be significantly larger than the ones on $\RHH$, we recommend to neglect the latter ones with respect to the former as a leading approximation.

In summary, if one uses Eq.~\ref{eq:cxNNLO} to calculate  $\sigmaHH$ we recommend to use the uncertainties from $\sigmaHHSMNNLO$ given in Ref.~\cite{MelladoGarcia:2150771} and assign no specific uncertainty to $\RHH$. Indeed, the uncertainties seem to be approximated well by the ones in the SM cross section prediction.

\subsection{Translation to the Higgs basis}
\label{sec:basis}

Obtaining the results of the fit in the H basis used by the LHCHXSWG2~\cite{Duehrssen-Debling:2001958} is straightforward. In Table~\ref{tab:higbasis} we provide the translation rules for the
coefficients of the operators we consider from our basis to the convention of the H basis.
In the latter, the coefficients of SM-like operators (the Higgs boson trilinear coupling and the top Yukawa interaction)
are parameterized as additive deviations from the SM values.
The pure BSM parameters in Eq.~\ref{eq:lag} are directly proportional to the corresponding
coefficients in the H basis. 

\begin{table}[h]
\centering
\small{
\begin{tabular}{ccc}
\toprule
Operator  & \multicolumn{2}{c}{Coefficient}  \\
($-\cal{ L}_{\PH}$) & our basis & H basis  \\
\midrule
$\PH^3$ &$\kappa_{\lambda}$ & $1 + \delta \lambda_3/\lambda_{\rm SM}$ \\[0.8mm]
$\frac{ m_t}{v} \PH\,( \bar{t_L}t_R + h.c.)$ & $\kappa_{t}$ & $1 + \delta y_t$ \\[0.8mm]
$\frac{ m_t}{v^2} \PH\PH ( \bar{t_L}t_R + h.c.)$ &$c_2$ & $y_t^{(2)}/2$ \\[0.8mm]
$-\frac{1}{4} \frac{\alpha_s}{3 \pi v} \PH\, G^{\mu \nu}G_{\mu\nu}$ & $c_g$ & $c_g^{\PH}\, 12 \pi^2$ \\[0.8mm]
$-\frac{1}{4} \frac{\alpha_s}{3 \pi v^2} \PH\PH G^{\mu \nu}G_{\mu\nu}$ &$c_{2g}$ & $-c_{gg}^{\PH}\, 12 \pi^2$ \\[0.8mm]
\bottomrule
\end{tabular}
}
\caption{\small
Translation of coefficients of operators (in terms of the physical fields) from our basis to the $\PH$ basis (neglecting $CP$ violating effects).
\label{tab:higbasis} }
\end{table}

The cross section ratio written in terms of the parameters in the H basis reads:

\begin{equation}
\begin{split}
R_{hh} = &
1
+ \delta\lambda_3 (A_{1}^{\PH} +  \delta\lambda_3\, A_{2}^{\PH})
+ \delta y_t \, (A_{3}^{\PH}  +   \delta y_t \, A_{4}\, )
+ \delta\lambda_3 \,\delta y_t\, A_{5}^{\PH} \\[2mm]
&
+ y_t^{(2)}\,( A_6^{\PH}  +  y_t^{(2)} \, A_7^{\PH} )
+  \delta\lambda_3 \, y_t^{(2)}  \, A_{8}^{\PH}
+  \delta y_t \, y_t^{(2)} \, A_{9}^{\PH}
+ c_{g}^{\PH}\,(A_{10}^{\PH}   +  c_{g}^{\PH}\, A_{11})
\\[2mm]
&
 + c_{gg}^{\PH}\, (A_{12}^{\PH}  +  c_{gg}^{\PH}\, A_{13}^{\PH})
 +  c_{g}^{\PH} c_{gg}^{\PH} \,A_{14}^{\PH}
 + \delta\lambda_3\,(
 c_{g}^{\PH} \,A_{15}^{\PH}
+    c_{gg}^{\PH} \, A_{16}^{\PH}
)
\\[2mm]
&
 + \delta y_t \,(   c_{g}^{\PH} \, A_{17}^{\PH}
+  c_{gg}^{\PH} \, A_{18}^{\PH}
)   +y_t^{(2)}(
 c_{g}^{\PH}\, A_{19}^{\PH}
+  c_{gg}^{\PH}\,  A_{20}^{\PH} )\,.
\end{split}
\label{eq:cxH}
\end{equation}

Although expanding the cross section up to quadratic order in the 
couplings leads to 20 coefficients $A_i^{\PH}$, clearly
the number of free coefficients in the fit remains the same as before. In fact, $\delta \lambda_3,\delta y_t$ always enter in the combination $(1 - \delta\, \cdot)^n$ in the matrix elements, relating different powers of the couplings. Connected to this, it is more convenient (and stable) to perform the fit actually in the parametrization of multiplicative deviations (Eq.~(\ref{eq:cx})), avoiding spurious coefficients.

The central values of the coefficients of Eq.~\ref{eq:cxH} are calculated using Table \ref{tab:higbasis} and shown in Table \ref{tab:coef_hbasis}.
The rescaling of the H-gluon contact interactions makes the corresponding coefficients typically much larger than the others.

\begin{table}[h]
\centering
\footnotesize{
\begin{tabular}{r r r r r r}
\toprule
$\sqrt{s} $ & 7\,TeV & 8\,TeV & 13\,TeV & 14\,TeV & 100\,TeV\\
\midrule
$A_{1}^H$ & -6.74 & -6.61 & -6.22 & -6.17 & -5.32 \\[0.8mm]
$A_{2}^H$ & 19.69 & 18.92 & 16.70 & 16.44 & 12.32 \\[0.8mm]
$A_{3}^H$ & 4.88 & 4.86 & 4.81 & 4.80 & 4.69 \\[0.8mm]
$A_{4}^H$ & 8.96 & 8.90 & 8.71 & 8.68 & 8.28 \\[0.8mm]
$A_{5}^H$ & -25.36 & -24.76 & -23.00 & -22.78 & -19.17 \\[0.8mm] 
$A_{6}^H$ & -2.84 & -2.84 & -2.84 & -2.85 & -2.9 \\[0.8mm]
$A_{7}^H$ & 2.46 & 2.47 & 2.54 & 2.55 & 2.89 \\[0.8mm] 
$A_{8}^H$ & 11.90 & 11.63 & 10.87 & 10.77 & 9.34 \\[0.8mm] 
$A_{9}^H$ & -7.22 & -7.19 & -7.10 & -7.09 & -7.02 \\[0.8mm] 
$A_{10}^H$ & -47.95 & -46.43 & -41.90 & -41.32 & -31.62 \\[0.8mm] 
$A_{11}^H$ & 1693.65 & 1621.64 & 1419.59 & 1396.29 & 1037.28 \\[0.8mm] 
$A_{12}^H$ & -138.90 & -134.89 & -121.47 & -119.61 & -76.31 \\[0.8mm] $A_{13}^H$ & 15922.80 & 16379.50 & 18682.60 & 19154.90 & 46060.00 \\[0.8mm]
$A_{14}^H$ & 8673.36 & 8475.37 & 7969.47 & 7916.62 & 7493.90 \\[0.8mm] $A_{15}^H$ & -17.38 & -20.98 & -29.47 & -30.16 & -35.19 \\[0.8mm] 
$A_{16}^H$ & 867.74 & 840.34 & 761.88 & 752.54 & 600.26 \\[0.8mm] 
$A_{17}^H$ & -141.60 & -136.56 & -121.87 & -120.05 & -90.28 \\[0.8mm] $A_{18}^H$ & -390.60 & -379.02 & -341.98 & -337.06 & -230.66 \\[0.8mm] $A_{19}^H$ & 97.58 & 94.69 & 86.32 & 85.34 & 216.14 \\[0.8mm] 
$A_{20}^H$ & 304.87 & 301.54 & 291.21 & 289.94 & 97.71 \\[0.8mm]
\bottomrule
\end{tabular}
}
\caption{Central values for the coefficients entering $R_{hh}$ (Eq. 
(\ref{eq:cxH})), employing the Higgs basis. \label{tab:coef_hbasis}}
\end{table}

\section{Conclusions}

In this document we have shown how the wide and high-dimensional space of
anomalous couplings that parametrize possible extensions of the standard model
can be investigated in a systematic way, employing the example of Higgs-pair production. 

We study the properties of 12 clusters, represented by benchmark points that describe the varying kinematic properties
of the full multi-dimensional phase-space, and we suggest a method to study how the upper limits on the cross section of a benchmark model derived by an experimental search can be extrapolated to the points of parameter space included in the corresponding cluster.

An analytical parametrization of the cross section valid for any point of the phase space is presented and shown to deliver a good approximation of the NNLO+NNLL prediction.
Precise uncertainties related to QCD parameters and missing oder effects are also 
offered. Using this information, an experimental analysis can easily perform an exhaustive scan of anomalous di-Higgs production
within the framework of the EFT.

\section{Postscript}

A recent result appeard after the end of this work~\cite{Borowka:2016ehy} indicates that the full top mass effects at NLO may have an impact dependent on $\mHH$ larger than the one predicted by the approximative calculations. The calculation performed in Ref.~\cite{Borowka:2016ehy} is assuing the SM case. Since no more generic calculations are yet availables for the BSM space under consideration, we let this interesting point for the further explorations of the clustering approach.

\section*{Acknowledgments}

We would like to thank Josh Bendavid and Olivier Bondu for precious help with the generators setup; Ken Mimasu for discussions and comments. 
We also thank  Amina Zghiche, Debdeep Ghosal and Serguei Ganjour for further cross checks.  A.C., F.G. and M.G. would like to express special thanks to the Mainz Institute for Theoretical Physics (MITP) for its hospitality. A.C. is  supported by  MIURFIRB RBFR12H1MW grant. M.D. is supported by grant CPDR155582 of Padua University. The research of F.G is supported by a Marie Curie Intra European Fellowship within the 7th European Community Framework Programme (grant no. PIEF-GA-2013-628224). 

\printbibliography

\end{document}